\documentclass[12pt]{iopart}

\usepackage{booktabs}
\usepackage{url}
\usepackage{graphicx}
\usepackage[compatibility=false]{caption} 
\usepackage[dvipsnames]{xcolor}
\expandafter\let\csname equation*\endcsname\relax
\expandafter\let\csname endequation*\endcsname\relax
\usepackage{amsmath}
\usepackage{amssymb}
\usepackage{bm}
\usepackage{cite} 

\newcommand{\dbp}{\delta \! B_{\parallel}}
\let\Pi\varPi

\makeatletter

\renewcommand\@makefnmark{\hbox{\textsuperscript{\normalfont\@thefnmark}}}
\renewcommand\@makefntext[1]{\parindent 1em\noindent\makebox[1.6em][l]{\footnotesize\rm\@thefnmark.\hfill}\footnotesize\rm #1}
\makeatother

\begin{document}

\title[On the transition to large fluxes in electromagnetic turbulence ]{On the transition to large fluxes and access to second stability in gyrokinetic simulations of electromagnetic turbulence in STEP}

\author{D. Kennedy$^{1,2}$, Y. Zhang$^{2,3,1}$, T. Adkins$^{4}$, P. G. Ivanov$^{1,2}$, F.J. Casson$^{1}$, H. G. Dudding$^{1}$, B. S. Patel$^{1}$, C. M. Roach$^1$, and H. R. Wilson$^{5}$}

\address{$^1$UKAEA (United Kingdom Atomic Energy Authority), Culham Campus, Abingdon,
Oxfordshire, OX14 3DB, UK}
\address{$^2$Rudolf Peierls Centre for Theoretical Physics, University of Oxford, Oxford, OX1 3JP, UK}
\address{$^3$Tokamak Energy Ltd, 173 Brook Drive, Milton Park, Abingdon, OX14 4SD, UK}
\address{$^4$Princeton Plasma Physics Laboratory, Princeton, New Jersey, 08540, USA}  
\address{$^5$UK Industrial Fusion Solutions Ltd., Abingdon, UK}

\ead{daniel.kennedy@ukaea.uk}

\begin{abstract}
This work investigates the nonlinear transition to large heat fluxes observed in local gyrokinetic simulations of electromagnetic turbulence in STEP~\cite{meyer2024}. Using the stress-balance framework of Zhang \textit{et al.}~\cite{zhang2026a,zhang2026b}, we confirm that the onset of extreme transport correlates with a critical value of $q^{2}\beta_{e}$, where $q$ is the safety factor and $\beta_{e}$ is the ratio of electron thermal pressure to magnetic pressure, and relate this to a limit on the poloidal beta $\beta_{\mathrm{pol}}$. Crucially, this critical value lies below any relevant linear stability limit in the ($q$, $\beta_{e})$ space (e.g., the onset of ideal or kinetic ballooning modes). 
 Using an extensive set of nonlinear gyrokinetic simulations, we demonstrate that the transition to large fluxes in STEP is governed by a balance between the electrostatic and magnetic-flutter stresses. We argue, and also show numerically, that larger-major-radius tokamaks reach the electromagnetic non-zonal regime at lower $\beta_e$, making this MHD-controlled saturation limit more accessible in reactor-scale devices than in small spherical tokamaks. We also demonstrate that access to a second-stable regime enables re-saturation at larger values of $\beta^{\prime}$. We further show that the ideal ballooning mode (IBM) threshold serves as a useful proxy for delineating this second-stable region and also as a qualitative guide to the onset of large fluxes. These results provide a predictive framework for identifying no-go zone predictions from local gyrokinetics and offer new insight into the electromagnetic saturation physics relevant to STEP and other high-$\beta_{e}$ devices.
\end{abstract}

%
%
%
%
%

\section{Introduction}
\label{sec:introduction}

The performance of magnetic-confinement-fusion (MCF) devices, such as spherical tokamaks (STs), is often limited by turbulent fluctuations, which dominate losses of heat, particles, and momentum. Predicting turbulence-driven transport is therefore essential for optimising future STs. The UK STEP Fusion programme~\cite{meyer2024,STEP,meyer2022} aims to develop a compact power plant generating over 100~MW of net electric power, based on the ST concept. In its first phase, the programme focuses on designing reference plasma equilibria using simplified core-transport models~\cite{tholerus2024}, as first-principles approaches remain too costly to apply on confinement or resistive timescales.

Reduced transport models have had notable success in conventional-aspect-ratio tokamaks~\cite{tglf,qualikiz,Staebler_2024,Bourdelle_2025}. However, STEP is expected to operate at significantly higher values of plasma $\beta_{e}$—the ratio of plasma (electron) pressure to magnetic pressure—than present-day machines~\cite{kaye2021}. In such regimes, electromagnetic (EM) effects play a leading role, the full dynamics of which reduced models often fail to capture. Due to these difficulties, local gyrokinetic (GK) simulations are increasingly relied upon to provide more accurate predictions in full-device modelling.

The first GK analysis of STEP’s preferred flat-top operating point, STEP-EC-HD~\cite{tholerus2024}, was conducted in~\cite{kennedy2023a} using the GK codes \texttt{GS2}~\cite{Rogers2000}, \texttt{GENE}~\cite{gene} and \texttt{CGYRO}~\cite{cgyro}. This study identified unstable \textit{hybrid} kinetic ballooning modes (hKBMs)—which combine features of kinetic ballooning modes (KBMs), ion-temperature-gradient-driven modes (ITGs), and trapped-electron modes (TEMs)—as the dominant instabilities, with subdominant microtearing modes (MTMs) also present (e.g.,~\cite{patel2021}). Subsequent nonlinear simulations~\cite{Giacomin2024,kennedy2024} revealed that, in the absence of equilibrium flow shear, hKBM-driven turbulence can generate particle and heat fluxes far exceeding the available heating and fuelling rates. Simulations of STEP's preferred flat-top operating point are often difficult to saturate without externally imposed equilibrium flow shear, with phases of apparent saturation punctuated by transient bursts of transport and a transition to extreme heat fluxes. However, it was also shown in \cite{Giacomin2024} that saturation at reasonable fluxes is possible at both larger and smaller values of $\beta_{e},$ assuming that the pressure gradient $\beta^{\prime}$ is scaled self-consistently.  

\begin{figure}
    \centering
    \includegraphics[]{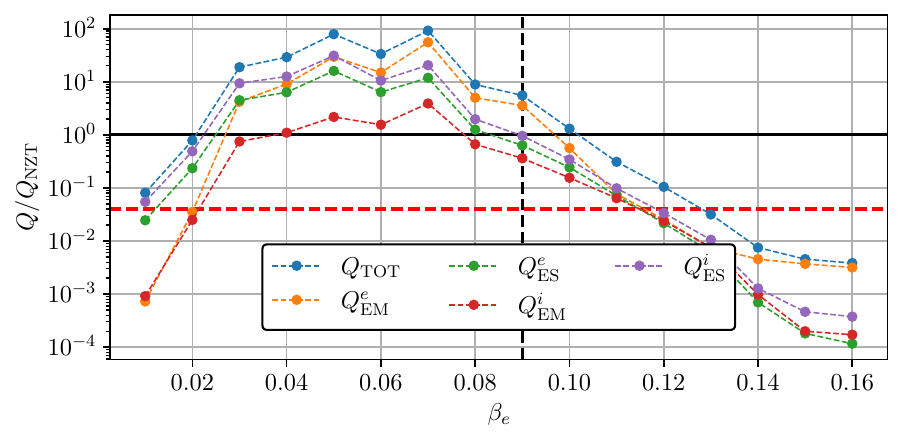}
    \caption{Electrostatic and electromagnetic electron and ion heat fluxes as functions of $\beta_{e}$ from nonlinear simulations using \texttt{GENE}. In this work, the pressure gradient, $\beta^{\prime},$ is always scaled self-consistently. All other parameters are kept fixed. Simulations are shown for a mid-radius surface in an early design of STEP-EC-HD examined in \cite{kennedy2023a,Giacomin2024}, with the nominal value of $\beta_{e}$ given by the black dashed line. The red horizontal line denotes the available heating power. All fluxes reported in this paper are normalised to $Q_{\mathrm{NZT}}$ (the non-zonal-transition flux, NZT) $= 50~\mathrm{MW\,m^{-2}}$.}
    \label{fig:step_betascan}
\end{figure}
\interfootnotelinepenalty=10000
Figure~\ref{fig:step_betascan} shows the turbulent heat flux calculated using local\footnote{All simulations in this paper are local, i.e., they are simulations performed in a domain of perpendicular size that is infinitesimal in comparison with the length scales over which the equilibrium varies. Gradients of the plasma equilibrium scale lengths are taken to be constant across the simulation domain and radial boundary conditions are assumed periodic.} gyrokinetic simulations at various values of $\beta_{e}$ in a more extensive scan using the gyrokinetic code \texttt{GENE}. In this work, the pressure gradient, $\beta^{\prime},$ is always scaled self-consistently so that $\beta^{\prime} \propto \beta_{e}$. Calculations are shown for a mid-radius surface for the STEP-EC-HD variant examined in \cite{kennedy2023a,Giacomin2024} and simulation parameters are identical to those used in \cite{Giacomin2024}. These simulations are performed in the absence of any equilibrium flow shear\footnote{STEP is expected to have minimal external momentum injection with the dominant plasma heating expected to be from a combination of ECCD (Electron Cyclotron Current Drive) and EBW (Electron Bernstein Wave) schemes~\cite{tholerus2024}.}. The black dashed line indicates the reference value of $\beta_{e}$ for this flux-surface in STEP-EC-HD. The red dashed line shows the heat flux that would correspond to the total plasma heating power crossing the surface. 

Two features are evident in Figure~\ref{fig:step_betascan}. Firstly, there exists a critical value of $\beta_e \in [0.02, 0.03]$ at which the total heat flux increases abruptly by approximately two orders of magnitude. As will be shown in Section~\ref{sec:numerical_simulations}, this transition is extremely sharp. Secondly, another, slower and less abrupt, transition occurs at $\beta_e > 0.10$, where the heat flux returns to comparatively low levels. The purpose of this paper is to analyse both of these transitions in detail. Whereas previous studies (e.g., \cite{Giacomin2024,Giacomin_Dickinson_Dorland_Mandell_Bokshi_Casson_Dudding_Kennedy_Patel_Roach_2025}) focused on quantitatively modelling turbulent transport in STEP scenarios, the present study instead aims to explore the nature of these transitions and to what extent we can use this insight to explore the structure of the parameter space accessible to STEP: identifying local gyrokinetic no-go zones, assessing viable operating points, and exploring access to and optimisation of such operating points.

\subsection{Electromagnetic high-transport states and the non-zonal transition}

The transitions identified in Figure~\ref{fig:step_betascan} are not specific to STEP-EC-HD. A transition to an EM high-transport state similar to that shown in Figure~\ref{fig:step_betascan} has long been observed in electromagnetic simulations of ITG-driven turbulence for the Cyclone Base Case (CBC)~\cite{Dimits2000} at finite $\beta_e$~\cite{Candy_2005,Pueschel2008,Pueschel2013}. In such scenarios, nonlinear simulations also frequently fail to saturate at the typically modest transport levels expected for electrostatic turbulence, and instead transition to a regime of very large heat fluxes. This is a nonlinear phenomenon, meaning that this transition is decoupled from the linear threshold of any instability, and is seemingly a Dimits-like transition \cite{Pueschel2014}. In this work, we use the term \textit{electromagnetic high-transport state} to refer to such cases where the system, after perhaps briefly appearing saturated, loses saturation and exhibits extreme transport. All statements regarding loss of saturation are based on finite-time simulations due to the computational cost of investigating such regimes, both in the present work and in previous studies. It cannot be excluded that the system would re-saturate if evolved for longer; no claim is made here concerning the true steady-state (long-time) behaviour.

Several mechanisms have been proposed to explain electromagnetic high-transport states (sometimes referred to as ``turbulence runaways'')  in local GK simulations. Early theories suggested that \emph{zonal flows} (ZFs), key regulators of electrostatic ITG turbulence~\cite{Diamond2005}, could be disrupted by tertiary EM instabilities~\cite{Waltz2010}, although subsequent studies showed that the amplitudes required for this mechanism are rarely reached in practice~\cite{Pueschel2013a}. An alternative explanation involves magnetic stochasticity, which becomes prominent at high $\beta_{e}$ and enables rapid electron transport along perturbed field lines~\cite{Nevins2011,Pueschel2013}. In this picture, when the perturbed magnetic field-line excursions exceed the zonal flow wavelength, rapid electron transport acts to ``short-circuit'' the zonal flows. Stochastic magnetic fields may also arise through the nonlinear excitation of subdominant microtearing modes~\cite{Hatch2013}. More recent work~\cite{Rath2022} attributes the transition to a near-cancellation between the electrostatic $\bm{E}\times \bm{B}$ Reynolds stress and the electromagnetic magnetic-flutter (Maxwell) stress, such that the net nonlinear drive of zonal flows is strongly reduced. This is consistent with a wide body of earlier work emphasising the role of stress balance in regulating zonal-flow dynamics in finite-$\beta$ and edge turbulence (e.g.,~\cite{Scott_2005,Abiteboul_2011,Xu_2006,Gurcan_2015,Dif-Pradalier_2009}). In this work, we find that in STEP-relevant electromagnetic turbulence the suppression of zonal-flow generation is not merely a consequence of the straightforward cancellation of nonlinear stresses, but of a systematic misalignment between the total stress and the zonal shear, leading to a negative stress-shear correlation that actively opposes flow amplification and indicates the need for a distinct electromagnetic turbulence saturation mechanism. This stress-balance picture is the gyrokinetic expression of a long-recognised feature of finite-$\beta$ turbulence: in the ideal-MHD limit of purely Alfv\'enic fluctuations the Reynolds and Maxwell stresses cancel exactly, and as $\beta_e$ increases this cancellation tightens and the net zonal-flow drive is progressively quenched~\cite{Diamond2005,Scott_2005}.

It is clear that zonal flows are key players in local GK turbulence and these previous findings raise fundamental questions about their role in nonlinear saturation. In particular, whilst zonal flow regulation of electrostatic ITG turbulence is now well understood~\cite{Dimits2000,Rogers2000,Lin1998,Ivanov2020,Ivanov2022}, the electromagnetic case remains less well developed. For example, whilst it is known that strong magnetic perturbations can disrupt zonal flows, enabling even subdominant EM instabilities such as KBMs~\cite{Mulholland2024} or hybrid modes~\cite{kennedy2023a,Giacomin2024,kennedy2024} to drive substantial transport, the presence of EM instabilities does not always correlate with high fluxes; in fact, some KBM simulations yield heat transport below the electrostatic limit~\cite{McKinney2021}. It has also been shown that zonal magnetic perturbations, rather than zonal flows, can play a key role in saturating MTM-driven turbulence in MAST-U~\cite{Giacomin2023microtearing}. More broadly, the theory of zonal-flow generation continues to advance, including the partition of the drive into beat-driven and spontaneous (modulational) contributions~\cite{Chen2024} and the role of phase-space zonal structures~\cite{ChenN2026}; we return to the relationship between these developments and the electromagnetic stress-balance picture studied here in Section~\ref{sec:conclusions}.

\begin{figure}
  \centering
  \includegraphics[width=\linewidth]{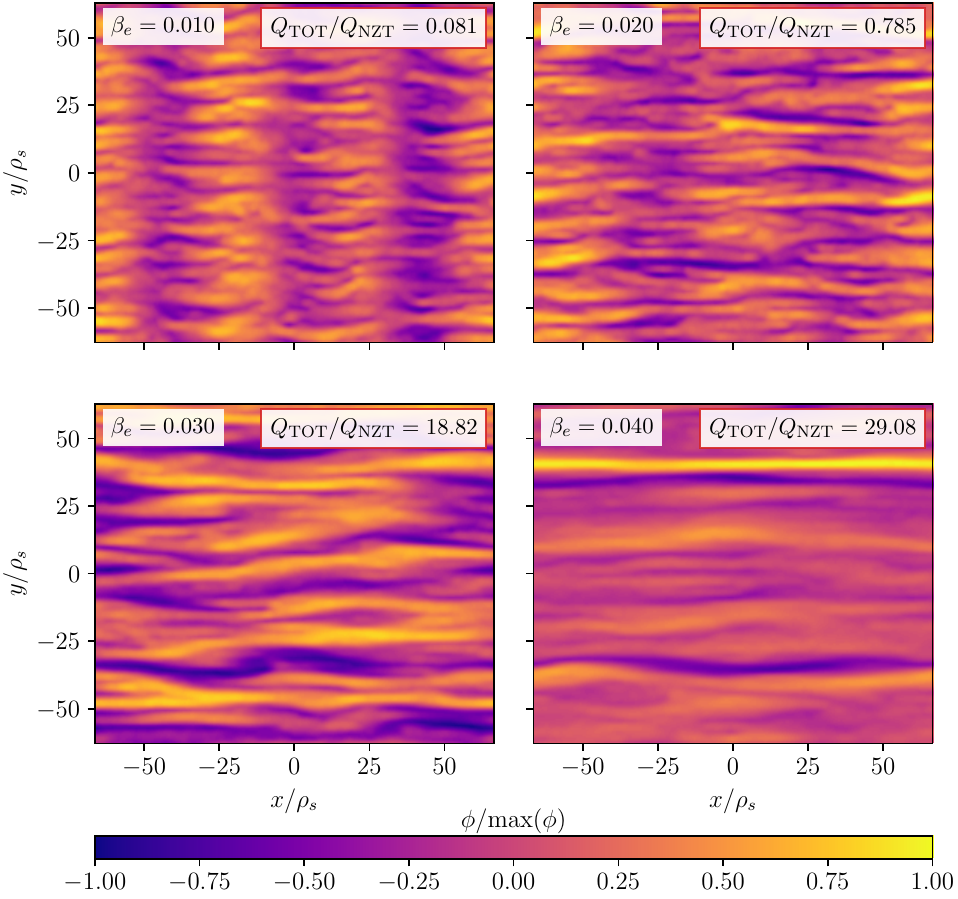}
\caption{
Snapshots of the time-averaged, perturbed electrostatic potential $\langle \phi(x, y) \rangle_t $ at the outboard midplane for four values of electron beta. Each panel is normalised such that $\phi \in [-1,1].$ The vertical banding at $\beta_e = 0.010$ and $\beta_{e} = 0.020$ (top row) indicates persistent zonal flows, while the more streamer-like horizontal structures at $\beta_e = 0.030$ and $\beta_{e} = 0.040$ (bottom row) suggest suppression of zonal flow activity leading to a transition to large fluxes. Time-averaged (over the last 10\% of the simulation) fluxes are reported in the top right-hand corner of each panel.
}
\label{fig:STEP_phi_contours}
\end{figure}

A defining feature of the EM high-transport regime is the appearance of streamers: coherent radial structures that transport heat efficiently across the domain~\cite{Waltz2010,Pueschel2013,Rath2022,Giacomin2024,kennedy2024}. Thus, following~\cite{Pueschel2008}, we refer to the first transition in Figure~\ref{fig:step_betascan}, from a simulation that saturates with moderate flux to one that enters an EM high-transport state, as the \emph{non-zonal transition}, indicating the suppression of zonal flow activity in the high-transport state. This transition is clearly visible in STEP simulations: as $\beta_e$ increases, time-averaged midplane snapshots of the electrostatic potential evolve from exhibiting clear vertically banded zonal flows, to become dominated by strong radially extended, streamer-like structures. This loss of zonal structure is illustrated in Figure~\ref{fig:STEP_phi_contours}, where the vertical coherence seen at lower $\beta_e$ gives way to broad, poloidally modulated streamers at higher $\beta_e.$ A key focus of this work is understanding the value of $\beta_{e}$ at which this transition occurs and what, if anything, can be done to moderate the impact of the transition.

Recent work by Zhang~\textit{et al.}~\cite{zhang2026a,zhang2026b} emphasises the importance of directly analysing zonal flow dynamics in nonlinear GK simulations. As discussed above, in the electrostatic case, saturation typically arises from self-organised zonal flows that shear apart turbulent eddies. In contrast, in the electromagnetic case, zonal flows are often suppressed or disrupted. Zhang~\textit{et al.}~\cite{zhang2026b} study the behaviour of zonal flows (and the sources of momentum that drive them) close to the transition. They find that the transition occurs when magnetic-flutter stress (the source of zonal flow from the EM part of the GK nonlinearity) extracts energy from zonal flows at a rate that exceeds the injection of energy from $\bm{E} \times \bm{B}$ stress, resulting in a regime with weak zonal flows and elevated heat fluxes. Zhang~\textit{et al.}~\cite{zhang2026b} were able to show numerically that this behaviour is observed across a range of configurations, including CBC~\cite{Dimits2000} and ST40~\cite{McNamara2024}. It was shown that in these geometries the non-zonal transition occurs when $q^2 \beta_e$ exceeds a critical threshold, $C_{\mathrm{nl}}$, below the threshold for ideal and kinetic ballooning mode instability. Here, and throughout, $q$ is the safety factor. The value of $C_{\mathrm{nl}}$ is not universal, but depends on the device and equilibrium shaping: for example, $C_{\mathrm{nl}} \approx 0.016$ for CBC and $C_{\mathrm{nl}} \approx 0.04$ for ST40. It was not clarified in~\cite{zhang2026b} which shaping factors determine this variation (see Sections~\ref{subsubsec:forward_transition_as_a_constraint_on_betapol} and~\ref{sec:ideal_ballooning_and_second_stabiity} of this paper).\footnote{We stress that $C_{\mathrm{nl}}$ is not a literal order-unity constant: beyond its dependence on device and flux-surface shaping, it is expected to depend on the magnetic shear $\hat{s}$ (as hinted at by the ideal-ballooning calculations of Section~\ref{sec:ideal_ballooning_and_second_stabiity}) and on any equilibrium $\bm{E}\times\bm{B}$ flow shear, neither of which is varied in the scans presented here.} Crucially, Zhang~\textit{et al.} also proposed a practical means of estimating the value of $C_{\mathrm{nl}}$ at which this transition occurs: showing that the ratio of magnetic-flutter to $\bm{E} \times \bm{B}$ stress scales approximately as $(q^2 \beta_e)^2$ at low $\beta_{e}$, enabling the non-zonal transition boundary to be estimated by extrapolating from a single nonlinear simulation.

In this paper, we examine and extend the stress-based interpretation of \cite{zhang2026a,zhang2026b}\footnote{At the time of writing, \cite{zhang2026b}, and the related work~\cite{zhang2026a}, are under consideration for publication; both are available as preprints (arXiv:2607.11789 and arXiv:2606.04616, respectively). The relevant results of~\cite{zhang2026a,zhang2026b} are summarised in a self-contained way in Section~\ref{sec:competition_of_stresses}, so that the present paper can be read independently of them.} and explore its relevance to STEP. In Section~\ref{sec:competition_of_stresses}, we briefly restate the main arguments of \cite{zhang2026b} and re-introduce the numerical diagnostics used therein. Whilst \cite{zhang2026b} focuses on the non-zonal transition and the competition between the $\bm{E} \times \bm{B}$ and magnetic-flutter components of the stress, here we emphasise and explore the contribution of the linear term and its role in accessing the low-flux state at larger values of $q^{2}\beta_{e}$. We also explore how the condition $q^2\beta_{e}$ can be recast as a limit on the poloidal beta and discuss this in the context of next-generation fusion power plants. In Section~\ref{sec:numerical_simulations} we perform nonlinear gyrokinetic simulations of the STEP-EC-HD flat-top using the diagnostics introduced in \cite{zhang2026b}. Our STEP-EC-HD simulations show excellent agreement between the observed transition threshold and the $q^{2}\beta_{e}=C_{\mathrm{nl}}$ criterion of~\cite{zhang2026a,zhang2026b}, providing the first gyrokinetic test of that criterion in reactor-relevant geometry. Using an extensive suite of over 100 nonlinear GK simulations in STEP-EC-HD and CBC geometry, we show that above the threshold in $q^2 \beta_e$, saturation at modest fluxes is achieved at higher pressure gradients than the values explored in \cite{zhang2026b}. In Section~\ref{sec:ideal_ballooning_and_second_stabiity}, we discuss the physics of the ideal ballooning mode (IBM) instability and why this mode seems to have some use as a proxy for the onset of both the non-zonal transition and the second transition identified in this paper. To be explicit: the stress-balance framework, the threshold $q^2\beta_e = C_{\mathrm{nl}}$, and the scaling $\Pi_{A_\parallel}/\Pi_\phi \propto (q^2\beta_e)^2$ are results of~\cite{zhang2026a,zhang2026b}, whereas the new contributions of the present work are (i) the first application and validation of that framework in STEP-EC-HD geometry, using an extensive suite of over $100$ nonlinear gyrokinetic simulations; (ii) an exploration of the role of $\dbp$ which is essential for GK modelling of STEP~\cite{kennedy2024}; (iii) the identification and analysis of the \emph{reverse} transition and the associated access to a second-stable regime at large $\beta'$, including the role of the linear stress $\Pi_{\mathrm{lin}}$; (iv) the recasting of the threshold as a constraint on the poloidal beta and its dependence on aspect ratio; and (v) the use of the ideal-ballooning-mode boundary as a practical proxy for both transitions. We give our conclusions in Section~\ref{sec:conclusions}.

\section{Zonal energy balance and the nonlinear $q^{2}\beta_{e}$ threshold}
\label{sec:competition_of_stresses}

Following~\cite{zhang2026a,zhang2026b}, we begin by summarising the coordinate conventions used therein. The equilibrium magnetic field is expressed in Clebsch form, $\bm{B} = \nabla \alpha \times \nabla \psi,$ with $\psi$ labelling flux surfaces (e.g., poloidal magnetic flux, set to zero on-axis) and $\alpha = \zeta - q(\psi)\theta$ labelling field lines within a surface \cite{Kruskal1958, DHaeseleer1991}. Here, $\theta$ is a poloidal angle coordinate parametrising position along a flux surface, and $\zeta$ is the toroidal angle. Working in the local, flux-tube approximation \cite{Beer1995}, we adopt field-aligned coordinates $(x, y, z)$, where
\begin{equation}
x = \frac{\mathrm{d}x}{\mathrm{d}\psi}\bigg\vert_{\psi_0}(\psi - \psi_0), \quad y = \frac{\mathrm{d}y}{\mathrm{d}\alpha}\bigg\vert_{\psi_0}(\alpha - \alpha_0)
\label{eq:flux_tube_coordinates}
\end{equation}
denote radial and binormal displacements, and $z$ is the coordinate along the fieldline. These are defined so that $\mathbf{B}/{B_0} = \nabla x \times \nabla y,$ with $x$ and $y$ having units of length and $B_0$ is a constant normalising magnetic field. Due to axisymmetry, equilibrium quantities are independent of $y.$ Any fluctuating quantity can thus be expanded as
\begin{equation}
g(\mathbf{r}) = \sum_{\mathbf{k}_\perp} g_{\mathbf{k}_\perp}(z)\, \mathrm{e}^{i(k_x x + k_y y)},
\end{equation}
since we assume statistical periodicity in the plane perpendicular to $\mathbf{B}$.

Given that zonal perturbations are, by definition, those that are constant on a flux-surface, these are naturally extracted from our system of equations by means of the flux-surface average, defined as
\begin{equation}
\langle \dots \rangle_\psi = \lim_{\Delta \psi \to 0} \frac{1}{\Delta V(\psi)} \left[ \int_{\Delta V(\psi)} (\dots) \, d^3r \middle/  \int_{\Delta V(\psi)} \, d^3r \right] ,
\label{eq:FSA}
\end{equation}
where $\Delta V(\psi) = V(\psi + \Delta \psi) - V(\psi)$ and $V(\psi)$ is the volume of the flux-surface labelled by $\psi$. 

In \cite{zhang2026b}, it was shown that the zonal fields in an electromagnetic plasma evolve according to \begin{equation}
\frac{\partial}{\partial t} \left\langle \sum_{\bm{k}_{\perp}} \mathrm{e}^{\mathrm{i}\bm{k}_{\perp} \cdot \bm{r}} \left[ \sum_{s} \frac{q_{s}^{2}n_{s}}{T_{s}} (1- \Gamma_{0s})\phi_{\bm{k}_{\perp}} - \sum_{s} q_s n_s \Gamma_{1s} \frac{\delta \! B_{\parallel \bm{k}_\perp}}{B}\right] \right\rangle_{\psi} = \Pi_{\mathrm{lin}} + \Pi_{\mathrm{turb}},
\label{eq:zonal_evolution_equation}.
\end{equation}
Here, the left-hand side represents the flux-surface-averaged time evolution of the generalised polarisation charge, including both electrostatic and magnetic compressibility effects. Equation~(\ref{eq:zonal_evolution_equation}) is the flux-surface-averaged charge-balance moment of the gyrokinetic--Maxwell system, with $\delta \! B_\parallel$ retained through the perpendicular piece of Amp\`{e}re's law\footnote{The closely related toroidal-angular-momentum balance, in which the same turbulent Reynolds and Maxwell stresses appear as radial fluxes of toroidal angular momentum, is derived from velocity moments of the gyrokinetic equation in the multiscale-gyrokinetic transport theory of~\cite{Abel2013}; that work also discusses the long-wavelength polarisation-charge interpretation of quasineutrality (its Appendix on gyrokinetic polarisation). The explicit zonal-charge evolution form used here, retaining $\delta \! B_\parallel$, is the one written down in~\cite{zhang2026b}.}. In this paper we are primarily interested in the zonal flow energy (associated with $\phi$). The right-hand side includes linear and nonlinear (turbulence-driven) sources of zonal flow. In Equation~(\ref{eq:zonal_evolution_equation}), $\phi$ is the perturbed electrostatic potential, $\delta \! B_\parallel$ is the fluctuation of the magnetic field parallel to its equilibrium direction, $\mathbf{b} = \bm{B}/B,$ and $h_s$ [which appears implicitly on the right-hand side of Equation~(\ref{eq:zonal_evolution_equation})] is the gyrotropic part of the perturbed guiding centre distribution function $\delta \! f_s = - (q_s \phi / {T_s}) F_s + h_s$ of species $s$ with equilibrium distribution $F_s = F_s(\psi)$, charge $q_s$, equilibrium density $n_s$, equilibrium temperature $T_s$, mass $m_s$, cyclotron frequency $\Omega_s = q_s B / m_s c$, thermal speed $v_{\mathrm{th}s} = \sqrt{{2 T_s}/{m_s}}$, and Larmor radius $\rho_s = {v_{\mathrm{th}s}}/{|\Omega_s|}$. The electromagnetic fields appearing in (\ref{eq:zonal_evolution_equation}), including the parallel component of the magnetic-vector potential $A_\parallel = \mathbf{b} \cdot \mathbf{A}$ (where $\mathbf{A}$ is the magnetic-vector potential), are determined by the quasineutrality condition, and by the parallel and perpendicular parts of Amp\`{e}re’s law, which are, respectively,
\begin{align}
\sum_s \frac{q_s^2 n_s}{T_s}\phi
&= \sum_s q_s \int \mathrm{d}^{3}\bm{v}\, \langle h_{s} \rangle_{\bm{r}},
\label{eq:quasineutrality} \\[0.5em]
\nabla^{2}_{\perp} A_\parallel
&= -\frac{4\pi}{c}\sum_s q_s \int \mathrm{d}^{3}\bm{v}\, v_\parallel \langle h_{s} \rangle_{\bm{r}},
\label{eq:ampere_parallel} \\[0.5em]
\nabla^{2}_{\perp} \delta \!B_\parallel
&= -\frac{4\pi}{c} \mathbf{b} \cdot \left[
\nabla_{\perp} \times \sum_s q_s \int \mathrm{d}^{3}\bm{v}\,
\langle \bm{v}_{\perp} h_{s} \rangle_{\bm{r}}
\right].
\label{eq:ampere_perpendicular}
\end{align}

Here, and throughout the remainder of this paper, $\langle \dots\rangle_{\bm{r}}$ and $\langle \dots \rangle_{\bm{R}}$ denote the standard gyroaverage (average over the gyroangle $\vartheta$) at constant real position $\bm{r}$ and guiding-centre position $\bm{R}_{s} = \bm{r} - \mathbf{b} \times \bm{v}_{\perp} /\Omega_{s}$, respectively, while
\begin{equation}
    \Gamma_{0s}(\alpha_s) = \mathrm{I}_{0}(\alpha_s) \mathrm{e}^{-\alpha_s}, \,\,\,\, \Gamma_{1s}(\alpha_s) = \left[\mathrm{I}_{0}(\alpha_s) - \mathrm{I}_1 (\alpha_s)\right] \mathrm{e}^{-\alpha_s}, \,\,\,\, \alpha_{s} = \frac{1}{2}k_{\perp}^{2}\rho_s^2,
\end{equation}
are functions of the perpendicular wavenumber $k_\perp = |\bm{k}_\perp|$, with $\bm{k}_\perp = k_x \nabla x + k_y \nabla y$, that capture finite-Larmor-radius effects (see, e.g., \cite{Howes2006}), and in which $\mathrm{I}_0$ and $\mathrm{I}_1$ are modified-Bessel functions of the first kind \cite{Abramowitz1972}. It is worth noting that the only assumption (beyond the standard GK ordering) employed in arriving at (\ref{eq:zonal_evolution_equation}) was that of the axisymmetry of the equilibrium, viz., (\ref{eq:zonal_evolution_equation}) does not depend on a particular choice of magnetic geometry, equilibrium plasma pressure profiles, plasma-beta, etc.

\subsection{Linear contribution to the zonal flow drive}
\label{subsec:linear_stress_definitions}

The first term on the right-hand-side of (\ref{eq:zonal_evolution_equation}) is the linear `stress'\footnote{Note that the terms appearing on the right-hand-side of (\ref{eq:zonal_evolution_equation}) are divergences of the radial current density, which is proportional to, but dimensionally different by a factor of B from, the stress.  
(\ref{eq:zonal_evolution_equation}) is therefore equivalent to the zonal momentum equation:
\begin{equation}
\frac{\partial}{\partial t}\,(\text{zonal momentum})
= \nabla \cdot \bm{\Pi}_{z},
\end{equation}
where $\bm{\Pi}_{z}$ is the total momentum flux (i.e.,\ the genuine 
stress tensor). In this sense, the right-hand-side contributions to (\ref{eq:zonal_evolution_equation}), 
$\Pi_{\mathrm{lin}}$ and $\Pi_{\mathrm{turb}}$
are strictly interpreted as the \emph{divergences of currents}, 
normalised according to the gyrokinetic conventions used here. 
For brevity, and following the established usage in the literature 
(e.g.,~\cite{Rath2022,zhang2026a}), we nevertheless refer to them simply as 
stresses throughout this work.}

\begin{equation}
\Pi_{\mathrm{lin}} \equiv - \left\langle \sum_s q_s \int\,  \mathrm{d}^3\bm{v} \,\, \left[ (\bm{v}_{ds} \cdot \nabla x) \left\langle \frac{\partial h_s}{\partial x} \right\rangle_{\bm{r}} \right] + \sum_{s^\prime}  \left\langle \left\langle C_{ss^\prime}^{(l)}[h_s] \right \rangle_{\bm{R}_s} \right\rangle_{\bm{r}} \right\rangle_{\psi},
\label{eq:linear_stress}
\end{equation}
that describes the (radial) drifting of the plasma across flux-surfaces due to the magnetic drifts contained within the drift velocity:
\begin{equation}
\mathbf{v}_{ds} = \frac{\mathbf{b}}{\Omega_s} \times \left[ v_\parallel^2 \mathbf{b} \cdot \nabla \mathbf{b} + \frac{1}{2}v_\perp^2 \nabla \log B \right],
\label{eq:magnetic_drifts}
\end{equation}
and the effects of inter-particle collisions on the gyrokinetic distribution function $h_s$ through the (linearised) collision operator $C_{ss^\prime}^{(l)}$. The form of (\ref{eq:linear_stress}) follows directly from the linear terms appearing in 
the gyrokinetic equation (e.g., \cite{Abel2013}) after taking the flux-surface average 
$\langle \cdots \rangle_{\psi}$: the averaging removes all contributions 
proportional to $\nabla_{\parallel}$ and $k_y$, so that only the terms proportional to radial 
derivatives survive. Moreover, because the 
background-gradient drive terms vanish after zonal averaging, 
$\Pi_{\mathrm{lin}}$ contains no free-energy drive, but only damping-like 
contributions from drifts and collisions. Physically, this reflects the 
fact that zonal modes evolve solely through radial redistribution and 
collisional relaxation (as well as nonlinear interactions), rather than being driven directly by linear instabilities. Given that core fusion plasmas are typically only weakly collisional~\cite{Abel2013}, we assume that the characteristic timescales associated with this collisional term are sufficiently long that it can be neglected in our analysis; we stress that no plasma is strictly collisionless, and that this assumption would need to be revisited in more collisional regimes (e.g., towards the edge or during ramp-up).

In \cite{zhang2026b} it was argued (and then verified numerically) that the dominant balance in setting the non-zonal transition threshold is determined by the nonlinear stresses (introduced in Section~\ref{subsec:nonlinear_stress_definitions}). However, here we point out the important role of the first term in the linear stress, which we later find to be crucial for obtaining the rollover of the fluxes observed in Figure~\ref{fig:step_betascan} at high pressure gradients, i.e., for allowing saturation at lower fluxes at $q^{2}\beta_{e}$ values that greatly exceed the non-zonal transition threshold (see Section~\ref{subsec:reverse_transition}). 

We can simplify (\ref{eq:linear_stress}) by appealing to gyrokinetic pressure balance (see e.g., \cite{Abel2013}) to give a useful identity (see e.g., Appendix C of \cite{kennedy2024}) relating the magnetic drift velocity to the pressure gradient viz, 
\begin{equation}
        \mathbf{v}_{ds} = \frac{\mathbf{b}}{\Omega_{s}} \times \left[ \left( v_{\parallel}^{2} + \frac{v_{\perp}^2 }{2} \right) (\mathbf{b}\cdot\nabla)\mathbf{b} - \frac{v_{\perp}^2 \mu_0 \nabla p}{2B^2} \right].
\end{equation}
Substituting this pressure-balanced form of the magnetic drift velocity into the first term of Equation~(\ref{eq:linear_stress}) gives
\begin{align}
\Pi_{\mathrm{lin}} &= - \left\langle \sum_s q_s \int \mathrm{d}^3\bm{v} \left[ \left( \frac{\mathbf{b}}{\Omega_s} \times \left[ \left( v_{\parallel}^{2} + \frac{v_{\perp}^2 }{2} \right) (\mathbf{b}\cdot\nabla)\mathbf{b}\right] \cdot \nabla x \right) \left\langle \frac{\partial h_s}{\partial x} \right\rangle_{\bm{r}} \right] \right\rangle_{\psi}.
\label{eq:Pi_lin_curv}
\end{align}
Note that the drift contribution from the pressure gradient has vanished since $\nabla p \propto \nabla x$. 

In Section~\ref{sec:ideal_ballooning_and_second_stabiity} we will argue that the stability of the ideal ballooning mode (IBM) contains similar physics to the stress-balance picture that we are building, with this physics entering through $\Pi_{\mathrm{lin}}$. A positive $\Pi_{\mathrm{lin}}$ acts as a local source in (\ref{eq:zonal_evolution_equation}) that \emph{boosts} the zonal momentum (recall, as noted above, that $\Pi_{\mathrm{lin}}$ is itself the divergence of the linear radial current); consistent with overall momentum conservation, any such local gain is compensated by a corresponding loss elsewhere in the radial profile. In the absence of collisions, $\Pi_{\mathrm{lin}}$ depends only on the magnetic drifts $\mathbf{v}_{ds}$ and the radial structure of the non-adiabatic distribution $h_s$, i.e.,\ on the equilibrium geometry (through parameters such as $q,$ the safety factor, and $\hat{s},$ the magnetic shear) and the radial form of the zonal response. 
$\Pi_{\mathrm{lin}}$ can change sign depending on phase relations, reflecting variations in how the radial redistribution from drifts couples to the zonal structure. The magnitude and character of $\Pi_{\mathrm{lin}}$ are controlled by the same geometric factors that set the linear stability of non-zonal modes, so $\Pi_{\mathrm{lin}}$ is directly linked to equilibrium geometry and its associated linear stability properties. These interplays are explored in Section~\ref{subsec:reverse_transition} and~\ref{app:linear_term}.
It is shown in~\ref{app:linear_term} that the curvature-driven linear stress scales as $r/qR,$ where $r$ is the local minor radius and $R$ is the major radius of the device, and is therefore enhanced and more geometrically sensitive in spherical tokamaks, while remaining finite in large-aspect-ratio devices. This implies that the same re-saturation mechanism can operate across aspect ratio but is pronounced in STEP-like plasmas.

\subsection{Nonlinear contribution to the zonal flow drive}
\label{subsec:nonlinear_stress_definitions}

The second term on the right-hand-side of (\ref{eq:zonal_evolution_equation}) is the turbulent (nonlinear) `stress'

\begin{equation}
\Pi_{\mathrm{turb}} \equiv - \left\langle \sum_s q_s \int \, \mathrm{d}^3\bm{v} \,\, \langle \bm{v}_\chi \cdot \nabla h_s \rangle_r \right\rangle_\psi, \label{eq:stress_turb}
\end{equation}
that describes the nonlinear advection of the plasma by the combination of effects contained within the drift velocity:
\begin{equation}
\bm{v}_\chi \equiv \frac{c}{B} \mathbf{b} \times \frac{\partial \langle \chi \rangle_{\bm{R}_s}}{\partial \bm{R}_{s}}, \quad \chi \equiv \phi - \frac{\mathbf{v} \cdot \mathbf{A}}{c}, 
\end{equation}
viz., the $\bm{E} \times \bm{B}$ drifts, parallel streaming along perturbed field lines, and the $\nabla B$ drift associated with perturbed magnetic-field. It will be useful to separate out each of these contributions by defining
\begin{align}
\Pi_\phi &\equiv - \left\langle \sum_s q_s \int \, \mathrm{d}^3\bm{v} \,\, \left\langle \left( \frac{c}{B} \mathbf{b} \times \frac{\partial}{\partial \bm{R}_{s}} \langle \phi \rangle_{\bm{R}_s} \right) \cdot \nabla h_s \right\rangle_{\bm{r}} \right\rangle_{\psi}, \label{eq:stress_ExB} \\
\Pi_{A_\parallel} &\equiv - \left\langle \sum_s q_s \int \, \mathrm{d}^3\bm{v} \,\, \left\langle \left( \frac{c}{B} \mathbf{b} \times \frac{\partial}{\partial \bm{R}_{s}} \left\langle -\frac{v_\parallel A_\parallel}{c} \right\rangle_{\bm{R}_s} \right) \cdot \nabla h_s \right\rangle_{\bm{r}} \right\rangle_{\psi}, \label{eq:stress_magnetic_flutter} \\
\Pi_{\delta \! B_\parallel} &\equiv - \left\langle \sum_s q_s \int \, \mathrm{d}^3\bm{v} \,\, \left\langle \left( \frac{c}{B} \mathbf{b} \times \frac{\partial}{\partial \bm{R}_{s}} \left\langle -\frac{\bm{v}_\perp \cdot \bm{A}_\perp}{c} \right\rangle_{\bm{R}_s} \right) \cdot \nabla h_s \right\rangle_{\bm{r}} \right\rangle_{\psi}. \label{eq:stress_compressive}
\end{align}
which we will henceforth refer to as the $\bm{E} \times \bm{B}$, magnetic-flutter, and $\dbp$ stresses, respectively. 

\subsection{Evolution of the zonal energy}
\label{subsec:evolution_of_zonal_energy}

Following \cite{zhang2026b}, we write (\ref{eq:zonal_evolution_equation}) 
as an energy equation describing the evolution of the zonal energy to render transparent how stresses either inject energy into or 
extract energy from the zonal vorticity, depending on the sign of the 
corresponding ZF torques\footnote{One can think of these torques as a turbulent zonal-flow viscosity.}. For brevity, we consider a plasma consisting of 
main ions with charge $Ze$ and electrons, and we neglect the $\delta \! B_\parallel$ contribution to the left-hand side of (\ref{eq:zonal_evolution_equation}), which we find numerically to be small (this approximation is derived and quantified in~\ref{app:zonal_energy_budget}). Furthermore, we focus on large-scale zonal
flows with $k_x \rho_i \ll 1$. In this long-wavelength limit, $\alpha_i \to 0$, and the electrostatic FLR factor may be expanded as $\Gamma_0(\alpha_i) \;\simeq\; 1 - \alpha_i + O(\alpha_i^2)$, so that the ion polarisation contribution reduces to
\begin{equation}
1 - \Gamma_0(\alpha_i) \;\simeq\; \alpha_i \;\simeq\; \tfrac{1}{2} k_\perp^2 \rho_i^2,
\end{equation}
meaning that the effective inertia of the zonal flow is reduced to a simple quadratic dependence on the radial wavenumber $k_{x}$, since for zonal flows $k_\perp = k_x\nabla x.$ 

Collecting terms, the zonal energy balance can be written in the compact form
\begin{equation}
\partial_{t} E_{\mathrm{ZF}}(k_{x}) \equiv \frac{\partial}{\partial t} \left( \frac{(Ze)^2 n_i k_{x}^{2} |\nabla x|^2 \rho_i^2}{4 T_i}
\left| \langle \phi \rangle_{\psi, k_x} \right|^2 \right)
=
T_{\mathrm{lin},k_x} + T_{\phi,k_x} + T_{A_\parallel,k_x} + T_{\delta B_\parallel,k_x}.
\label{eq:zonal_energy_balance}
\end{equation}
The term on the left-hand side represents the polarisation energy stored
in the zonal potential\footnote{Note that (\ref{eq:zonal_energy_balance}) holds in this form only in the large-aspect-ratio limit, where the geometric factors entering the polarisation term can be taken outside the flux-surface average on the left-hand side of (\ref{eq:zonal_evolution_equation}). Nevertheless, as shown in~\cite{zhang2026b}, this expression appears sufficient for the purposes of the present discussion. A more general definition of zonal-flow energy in realistic magnetic geometry is left for future work.}. The terms on the right-hand side are the corresponding ``energy transfers'', which act to inject or extract energy from the zonal 
component.

In (\ref{eq:zonal_energy_balance}) we have defined the ZF torques:
\begin{align}
T_{\mathrm{lin},k_x} &= \mathrm{Re}\left( \langle \phi \rangle^*_{\psi,k_x} \, \Pi_{\mathrm{lin},k_x} \right), \label{eq:lin_transfer} \\
T_{\phi,k_x} &= \mathrm{Re}\left( \langle \phi \rangle^*_{\psi,k_x} \, \Pi_{\phi,k_x} \right), \label{eq:phi_transfer} \\
T_{A_\parallel,k_x}  &= \mathrm{Re}\left( \langle \phi \rangle^*_{\psi,k_x} \, \Pi_{A_\parallel,k_x} \right),
\label{eq:apar_transfer} \\ 
T_{\delta\!B_\parallel,k_x} &= \mathrm{Re}\left( \langle \phi \rangle^*_{\psi,k_x} \, \Pi_{\delta\!B_\parallel,k_x} \right)
\label{eq:bpar_transfer}. 
\end{align}

Note that the $\delta\!B_\parallel$ part of the generalised polarisation that we have dropped here would augment the right-hand side of (\ref{eq:zonal_energy_balance}) by a single exchange term; this term, and the compressive zonal energy $E_{\delta\!B_\parallel}$, are derived from the full zonal vorticity equation and found, directly from the simulations, to be small across the entire $(q,\beta_e)$ parameter space in~\ref{app:zonal_energy_budget}.

In well-developed electromagnetic turbulence, such as that seen in gyrokinetic simulations of STEP-relevant plasmas, it is often the case that turbulent amplitudes grow large enough that the dynamics are dominated by nonlinear interactions. This regime is particularly evident during the electromagnetic high-transport state, where the heat flux grows rapidly in time \cite{Ivanov2020, Ivanov2022,Giacomin2024,kennedy2024}. As argued in \cite{zhang2026b}, in such cases, the leading-order balance in (\ref{eq:zonal_evolution_equation}) is between the time-derivative on the left-hand side of Equation~(\ref{eq:zonal_energy_balance}) and the turbulent stress terms on the right-hand side of Equation~(\ref{eq:zonal_energy_balance}). Consequently, the sign and temporal persistence of $T_{\mathrm{turb}}=T_{\phi,k_x}+T_{A_\parallel,k_x}+T_{\delta\!B_\parallel,k_x},$  provides a criterion for zonal saturation. If $T_{\mathrm{turb}}$ is positive and boosts ZF over many nonlinear turnover times, zonal flows are sustained. If, instead, $T_{\mathrm{turb}}$ is negative and damps the ZFs over many nonlinear turnover times, zonal flows are suppressed, giving way to streamer-dominated states with elevated heat fluxes.

Zhang \textit{et al.}~\cite{zhang2026a,zhang2026b}  quantified this transition by analysing the relative contributions of (\ref{eq:stress_ExB})–(\ref{eq:stress_compressive}) and showed that non-zonal states emerge when the energy extracted from zonal flows via magnetic stresses, particularly $\Pi_{A_\parallel},$ matches or exceeds the energy injection via $\Pi_\phi$\footnote{It is possible to show that, in the long-wavelength limit, $\Pi_\phi$ can be decomposed into Reynolds and diamagnetic stress contributions. The Reynolds stress has, on average, the same sign as the zonal shear and therefore reinforces the zonal flows that generate the shear. The sign of the diamagnetic stress cannot be determined \textit{a priori}, though it often opposes the Reynolds stress; previous work~\cite{Ivanov2020,Ivanov2022} has shown that the Dimits transition~\cite{Dimits2000} in a simple \(Z\)-pinch geometry is governed by whether the diamagnetic stress is able to overcome the Reynolds stress and thereby destroy the zonal flows responsible for reduced transport. By contrast, the Maxwell stress always has the opposite sign to the zonal shear and therefore acts to weaken zonal flows. In the electromagnetic regime, this reduces the net turbulent drive and makes the maintenance of zonal flows more difficult.}. Furthermore, and of particular importance for the design of next-generation plasma equilibria, they identified a practical criterion for this transition: $q^2\beta_{e} = C_{\mathrm{nl}},$ with $C_{\mathrm{nl}}$ empirically found to satisfy $C_{\mathrm{nl}} < C_{\mathrm{KBM}} < C_{\mathrm{IBM}}$ where $C_{\mathrm{KBM}}$ is the value at which the Kinetic Ballooning Mode (KBM) becomes linearly unstable (subdominantly) and $C_{\mathrm{IBM}}$ is the value at which the Ideal Ballooning Mode (IBM) becomes linearly unstable (see Section~\ref{sec:ideal_ballooning_and_second_stabiity}). This result implies that nonlinear electromagnetic turbulence can result in extreme transport, even when the plasma remains linearly stable to KBMs (as also seen in \cite{Pueschel2013}). \cite{zhang2026b} also proposed an empirical scaling law, $\Pi_{A_{\parallel}}/\Pi_{\phi} \propto (q^{2}\beta_{e})^{2},$ allowing the critical value of beta for the onset of large fluxes to be estimated from a single simulation at low-$\beta_{e}$. This framework provides a predictive theory of electromagnetic saturation, accounting for the interplay between the nonlinear stresses. 

\subsection{Threshold for the non-zonal transition in a tokamak} \label{sec:tokamak_threshold}

In \cite{zhang2026a,zhang2026b}, the authors argue that the critical plasma beta, $q^{2}\beta_{e,\mathrm{crit}} = C_{\mathrm{nl}},$ above which electromagnetic turbulence will fail to saturate, arises from the competition between electrostatic and electromagnetic (Alfv\'{e}nic) dynamics. The quantity $q^{2}\beta_e$ is the gyrokinetic analogue of the finite-$\beta$ parameter $\hat\beta=\beta(qR/L_n)^2/2$, which measures the ratio of the drift to shear-Alfv\'en frequency (see e.g.,~\cite{Diamond2005}). The Alfv\'{e}n speed scales as $v_A \sim v_{\mathrm{thi}}/\sqrt{\beta}$ (assuming $T_e = T_i$). At low $\beta_{e},$ the Alfv\'{e}nic propagation time along the field,
$\tau_A \sim (k_\parallel v_A)^{-1}$, is therefore short compared to the characteristic timescale of drift-wave and zonal-flow dynamics. In this regime, inductive magnetic perturbations respond rapidly to electrostatic forcing through parallel force balance. As a result, the magnetic-flutter contribution to the zonal-flow stress is parametrically small, and saturation is governed primarily by $\bm{E}\times\bm{B}$ dynamics. 
 However, as $ \beta_e $ increases, $ v_A $ decreases relative to $v_{\mathrm{thi}}$ and electromagnetic effects become dynamically relevant. Asymptotically, this will happen when the Alfv\'{e}n frequency for the lowest parallel wavenumber mode, $\omega_A^{\min} \sim  k_\parallel^{\rm min} v_A \sim v_A / (qR)$, becomes comparable to the MHD interchange growth rate $ \gamma_{\text{MHD}} \sim v_{\mathrm{thi}} / \sqrt{RL_{T_i}}$. Equating these two timescales yields a critical beta,
\begin{equation}
\beta_{e,\mathrm{crit}} \sim \frac{L_{T_i}}{q^2 R}.
\label{eq:critical_ion_beta}
\end{equation}
This marks the threshold above which Alfv\'{e}nic perturbations can interact strongly with drift waves, having comparable timescales, allowing magnetic flutter and its associated stresses to influence the turbulence\footnote{The estimate~(\ref{eq:critical_ion_beta}) can also be derived directly from quasilinear estimates of the stresses~\cite{zhang2026b}.}. 

\subsubsection{Critical value of beta poloidal}

Equation~(\ref{eq:critical_ion_beta}) for the critical beta
can be re-expressed in terms of the poloidal beta, a more geometric quantity that characterises the pressure-induced distortion of magnetic field lines. The poloidal beta is defined as (e.g., \cite{wesson2011tokamaks})
\begin{equation}
\beta_{\mathrm{pol}} = \frac{2\mu_{0} \langle p \rangle}{B_\theta^2},
\end{equation}
where \( \langle \cdots \rangle \) is the average over a flux surface, and \( B_\theta \) is the characteristic poloidal magnetic field strength. We can thus rewrite (\ref{eq:critical_ion_beta}) as
\begin{equation}
\beta_{\mathrm{pol,crit}}
\sim
\beta_{e,\mathrm{crit}}
\frac{B^2}{\langle B_\theta^2\rangle}
\sim
\frac{L_{T_i}}{q^2 R}
\frac{B^2}{\langle B_\theta^2\rangle}.
\label{eq:beta_pol_crit_general}
\end{equation}

Using the standard tokamak estimate $B_\theta/B\sim\epsilon/q$ (with $\epsilon = r/R$ the inverse aspect ratio), so that $B^2/\langle B_\theta^2\rangle\sim(q/\epsilon)^2,$ the threshold can be written compactly as
\begin{equation}
q^2\beta_{e,\mathrm{crit}}\;\sim\;\epsilon^2\,\beta_{\mathrm{pol,crit}},
\qquad\text{equivalently}\qquad
\beta_{\mathrm{pol,crit}}\;\sim\;\frac{C_{\mathrm{nl}}}{\epsilon^2}.
\label{eq:q2beta_betapol_main}
\end{equation}
The criterion $q^2\beta_e = C_{\mathrm{nl}}$ is therefore equivalently a limit on the poloidal beta, $\beta_{\mathrm{pol}} = C_{\mathrm{nl}}/\epsilon^2.$ The explicit factor of $\epsilon^2$ also clarifies why the calibrated constant $C_{\mathrm{nl}}\sim\epsilon^2\,\beta_{\mathrm{pol,crit}}$ is device-specific rather than universal: at fixed critical poloidal beta, a tighter aspect ratio (larger $\epsilon$) returns a larger $C_{\mathrm{nl}}.$

\subsubsection{Electromagnetic transition in large-aspect-ratio tokamaks}

An important consequence of the electromagnetic high-transport state being connected to Alfv\'enic dynamics is that the transition is more readily accessed in large-aspect-ratio devices. 
The Alfv\'en frequency, $\omega_A = k_\parallel v_A$, represents the restoring rate due to magnetic tension along field lines. 
In large-aspect-ratio geometry the parallel connection length scales as $L_\parallel \sim qR$, where $R$ is the major radius of the flux surface, so the lowest parallel wavenumber is $k_\parallel^{\min} \sim 1/(qR)$. Increasing \emph{either} $q$ or $R$ therefore lengthens the field lines and lowers $k_\parallel^{\min}$, weakening the magnetic tension; at fixed $q$ the relevant dependence is on $R$.
Consequently, Alfv\'enic dynamics are more easily excited and the transition occurs at lower $\beta_e$ in large-aspect-ratio configurations.

This provides a natural explanation for why the non-zonal transition occurs at substantially lower $\beta_e$ (at comparable $q$) in the large-aspect-ratio CBC than in the tight-aspect-ratio geometry of ST40, as reported in \cite{zhang2026b}. 
The scaling of the threshold with the flux-surface major radius $R$ is examined further in Section~\ref{subsubsec:forward_transition_as_a_constraint_on_betapol}.

\subsection{Connection to MHD and the ideal ballooning mode}

The critical condition $q^{2}\beta_{e,\mathrm{crit}}= C_{\mathrm{nl}}$ that marks the onset of the non-zonal transition is not a peculiarity of gyrokinetic turbulence, but rather appears to be the nonlinear manifestation of a familiar reduced-MHD force balance. It is the gyrokinetic counterpart of the ``Alfv\'enisation'' of finite-$\beta$ turbulence (see e.g.,~\cite{Diamond2005}), with $q^2\beta_e$ playing the role of $\hat\beta$. The threshold for the non-zonal transition corresponds to the point at which field-line bending by Alfv\'enic currents can no longer instantaneously neutralise pressure-driven curvature currents, i.e., when the Alfv\'en frequency at the lowest parallel wavenumber, $k_{\parallel}^{\min} v_A$, becomes comparable to the MHD interchange (or ballooning) growth rate. In this regime, the magnetic field loses its effective stiffness, allowing pressure gradients to drive dynamically significant magnetic perturbations. Importantly, the plasma may remain linearly stable to ideal-MHD ballooning modes; the threshold instead marks a transition from an adiabatic, field-line-tied electromagnetic response to one in which Alfv\'enic dynamics acquire inertia and slow to the point where it can couple to the electrostatic turbulence and grow. In MHD language this corresponds to crossing a critical poloidal beta, which measures the ratio of pressure to poloidal field-line tension. Gyrokinetics does not alter this geometric stability boundary, but it determines how it is realised nonlinearly: once field-line bending becomes slow, magnetic flutter and its associated Maxwell stresses couple efficiently into the zonal-flow momentum balance, undermining electrostatic saturation and enabling the transition to a non-zonal, high-transport state. This connection to the MHD ballooning mode is explored further in Section~\ref{sec:ideal_ballooning_and_second_stabiity}.

We begin by testing whether the stress-balance picture of~\cite{zhang2026a,zhang2026b} also accounts for the non-zonal transition we observe in STEP-EC-HD.

\section{Numerical Simulations} \label{sec:numerical_simulations}

The numerical simulations in this paper focus primarily on a single equilibrium flux-surface taken from close to mid-radius ($q = 3.5, \Psi_n = 0.49$) in the STEP reference scenario STEP-EC-HD\footnote{SimDB UUID: 2bb77572-d832-11ec-b2e3-679f5f37cafe, Alias: smars/jetto/step/88888/apr2922/seq-1} (where EC stands for Electron Cyclotron heating and current drive, and HD stands for High Density). This is an early iteration of the STEP design concept which was designed to deliver a fusion power $P_{\mathrm{fus}} = 1.8$~GW. We note that this is precisely the same flux surface as examined in \cite{kennedy2023a, Giacomin2024,kennedy2024}, a choice that has been made intentionally to allow our results to be compared with those reported elsewhere. We take this flux-surface as a starting point on which any parameter scans are based. 

The GK analysis presented in this work was facilitated by \texttt{pyrokinetics}~\cite{pyrokinetics,pyrokinetics2024}, a Python library developed to facilitate pre- and post-processing of gyrokinetic analysis performed using a range of different GK codes. \texttt{pyrokinetics} can be used to build GK inputs from experimental data, and also contains an ideal-ballooning solver, based on \cite{Gaur_Buller_Ruth_Landreman_Abel_Dorland_2023}, which has been used throughout this work (see Section~\ref{sec:ideal_ballooning_and_second_stabiity}). For the simulations reported here, we used \texttt{pyrokinetics} to generate a Miller parameterisation \cite{miller1998} of the local magnetic equilibrium geometry, and the shaping parameters
were fitted to the chosen surface. Table \ref{tab:q35_step_ec_hd_final} provides the values of various local equilibrium quantities at the flux surface examined in this paper, including magnetic shear $\hat{s}$, safety factor $q$, normalised minor radius $r/a$, elongation $\kappa$ and its radial derivative $\kappa^\prime$, triangularity $\delta$ and its radial derivative $\delta^\prime$, the radial derivative of the Shafranov shift $\Delta^\prime$, and the normalised inverse density and temperature gradient scale lengths of species $s$, $a/L_{ns}$ and $a/L_{Ts},$ respectively. Our simulations evolve two species: electrons and a deuterium-tritium mix, and neglect entirely impurities and fast particles. The interested reader is referred to \cite{tholerus2024,kennedy2023a,Giacomin2024} for more details on the equilibrium and to \cite{Bokshi2025} for information on impurity and fast particle modelling in STEP.  

\begin{table}[htbp]
    \centering
    \caption{
    Local parameters for the $q = 3.5$ surface in the STEP-EC-HD configuration. The binormal wavenumber $k_y^{n=1} \rho_s$ corresponds to toroidal mode number $n = 1$.
    }
    \label{tab:q35_step_ec_hd_final}
    \begin{tabular}{l@{\hskip 0.8cm}r@{\hskip 1.2cm}l@{\hskip 0.8cm}r}
        \toprule
        Parameter & Value & Parameter & Value \\
        \midrule
        $\beta_e$             & 0.09    & $\beta'$              & $-0.48$ \\
        $q$                   & 3.5     & $\hat{s}$             & 1.20 \\
        $\Psi_n$              & 0.49    & $r/a$          & 0.64 \\
        $\kappa$              & 2.56    & $\kappa'$             & 0.06 \\
        $\delta$              & 0.29    & $\delta'$             & 0.46 \\
        $\Delta'$             & $-0.40$ & $k_y^{n=1} \rho_s$    & 0.0047 \\
        $a/L_{n_e}$           & 1.06    & $a/L_{T_e}$           & 1.58 \\
        $a/L_{n_i}$           & 1.06    & $a/L_{T_i}$           & 1.82 \\
        \bottomrule
    \end{tabular}
\end{table}

In this work, we use the GK code \texttt{stella}~\cite{Barnes2019} and the stress diagnostics introduced in \cite{zhang2026b} to quantitatively study the zonal-flow dynamics across the transition boundary. \texttt{stella} calculates the gyrokinetic nonlinearity partitioned by field [i.e., it calculates and returns the terms appearing inside the angular brackets in Equations (\ref{eq:stress_ExB}) --(\ref{eq:stress_compressive})], as functions of $(k_{x}, k_{y}, z, t),$ via additional evaluations of the Poisson bracket. In addition, \texttt{stella} outputs the time-resolved $\partial \langle \phi\rangle_{\psi}/\partial t$ and $\partial \langle \dbp\rangle_{\psi}/\partial t,$ which are then used to construct the terms on the left-hand side of Equation~(\ref{eq:zonal_evolution_equation}). $\Pi_{\mathrm{lin}}$ is calculated from the residual of~(\ref{eq:zonal_evolution_equation}). These terms are then used to calculate the transfer functions~(\ref{eq:lin_transfer})-(\ref{eq:bpar_transfer}), which are now functions of $k_{x}$ and $t.$ It was shown in \cite{zhang2026b} that the dominant contribution to these terms comes from the largest scales in the system (the intuition for this is that the scales at which the zonal flows are most effective at suppressing turbulence ought to be larger than the outer scale of the turbulence). Thus, we define the large-scale transfers / ZF torques as, e.g.,
\begin{equation}
T_\phi = \sum_{|k_x| < k_{y0}} T_{\phi,k_x},
\label{eq:cutoff}
\end{equation}
where \( k_{y0} \) is the outer scale of the turbulence, which we define as the wavenumber corresponding to the peak of the heat flux spectrum in \( k_y \). Since drift-wave turbulence is typically anisotropic with $k_{x0} \lesssim k_{y0}$, this choice ensures that all large-scale radial modes are included in the sum (i.e., this is typically a conservative choice).

\texttt{stella} uses velocity space coordinates $(v_\parallel, \mu),$ the velocity component parallel to the field direction and the magnetic moment. This is identical to the discretisation used in \texttt{GENE} allowing direct comparison between the two. Over 100 nonlinear simulations were performed for the results in this section. The typical grid size used for STEP calculations in the parallel, binormal, radial, energy and pitch-angle dimensions, is given in Table~\ref{tab:resolution}. The grid size is comparable to those used in previous studies of the STEP-EC-HD equilibrium~\cite{Giacomin2024,kennedy2024}\footnote{It was shown in \cite{kennedy2024} that the salient nonlinear physics in GK simulations of STEP-EC-HD can actually be recovered evolving far fewer modes than evolved in the most highly-resolved STEP-EC-HD simulations~\cite{Giacomin2024}. The grids used in this work are a compromise between \cite{Giacomin2024} and \cite{kennedy2024}.}. \texttt{stella} simulations used the Dougherty collision operator (e.g., \cite{Barnes2019}) to model inter-particle collisions. In nonlinear simulations the radial box size $L_x = j/(\hat{s} k_{y,\mathrm{min}})$ is set to accommodate exactly eight rational surfaces of the lowest finite toroidal mode number by choosing the integer $j=8$, motivated by the numerical convergence study in~\cite{Giacomin2024}. Each simulation was run for over 96 hours of wallclock time using 16 CPU-only blades on the Data Centric General Purpose (DCGP) partition of the supercomputing system, PITAGORA, with each blade comprising two AMD Turin 128-core processors. These simulations are computationally expensive since additional computations of the GK nonlinearity (the standard nonlinear term with additional velocity space integrals and sums over species of separated elements) are required to calculate the stresses.


\begin{table}[h!]
\caption{Nominal resolution parameters for linear (L) and nonlinear (NL) simulations. Here $n_{k_x}$ is the spectral radial resolution, $n_{k_y}$ is the number of binormal Fourier modes, $n_z$ is the number of grid points along the field line, $n_{v_\parallel}$ and $n_\mu$ are the velocity-space resolutions, and $k_{y}^{\mathrm{min}}$ $(k_{x}^{\mathrm{min}})$ the wavenumber of the smallest binormal (radial) wavenumber (other than the zonal mode) evolved in the simulation. {The radial box size includes 1 or 8 rational surfaces in linear and nonlinear calculations, respectively.}}
\label{tab:resolution}
\centering
\renewcommand{\arraystretch}{1.5}
\begin{tabular}{@{}lcccccccccc@{}}
\toprule
\textbf{Simulation} & \( n_{k_x} \) & \(n_{k_y} \) & \( n_{z} \) & \( n_{v_\parallel} \) & \( n_{\mu} \) & \( k_{x}^{\mathrm{min}} \) & \( k_{y}^{\mathrm{min}} \) \\ \midrule
L           & 5          & 1           & 64                    & 32    & 16           & -             & -                      \\
NL          & 128          & 32              & 32                    & 32    & 16                & 0.025                     & 0.025  \\
 \bottomrule
\end{tabular}
\end{table}

Figure~\ref{fig:q-beta-step} summarises the results of a parameter scan over the $(q, \beta_e)$ plane using over 100 nonlinear gyrokinetic simulations of the STEP-EC-HD equilibrium performed with \texttt{stella}. Simulations that converge to a saturated state with a heat flux $Q_{\mathrm{TOT}} < Q_{\mathrm{NZT}} \equiv 50~\mathrm{MW\,m^{-2}}$ are marked by blue circles. Simulations which fail to saturate within the simulated timescale are marked by red crosses. A simulation is therefore marked with a red cross if either (i) the heat-flux time series over the final 10\% of the run fails an Augmented Dickey–Fuller stationarity test (at the 5\% level, $p > 0.05,$ see discussion in~\cite{Giacomin2024}), i.e., the flux remains statistically non-stationary and hence shows no evidence of saturation by the end of the run, or (ii) the final value of $Q_{\mathrm{TOT}}$ exceeds $Q_{\mathrm{NZT}}$ with relative fluctuations below 10\% over the final 10\% of the time series, indicating convergence to a high-transport \interfootnotelinepenalty=10000 state.\protect{\footnote{The criterion of $Q_{\mathrm{NZT}} \equiv 50~\mathrm{MW\,m^{-2}}$ is a somewhat arbitrary cut-off chosen to represent an increase in the flux by an order of magnitude from a low-$\beta_{e}$ well saturated state. It will be shown that across the transition the increase in the heat flux is very large; i.e., this can be thought of as a threshold in $\mathrm{d}Q_{\mathrm{TOT}}/\mathrm{d}\beta_{e}.$ A red cross therefore records the absence of evidence of saturation within the simulated time rather than a specific flux level; a borderline simulation just inside the boundary could in principle relax to a saturated state given substantially longer integration, but because the transition is a jump of one to two orders of magnitude across a narrow window in $\beta_{e}$ this would not move the location of the boundary.} Note that the transition is a jump of one to two orders of magnitude in $Q_{\mathrm{TOT}}$ across a narrow window in $\beta_e$ (Section~\ref{subsec:forward_transition}), and because $Q_{\mathrm{NZT}}=50~\mathrm{MW\,m^{-2}}$ already greatly exceeds the total plasma heating power crossing the surface, the blue/red classification in Figure~\ref{fig:q-beta-step} is essentially independent of the precise value of $Q_{\mathrm{NZT}}$: varying it by a factor of a few moves no points across the boundary. The black dashed line in Figure~\ref{fig:q-beta-step} marks the ideal ballooning mode (IBM) threshold computed from the ideal ballooning solver implemented in \texttt{pyrokinetics} (see discussion in Section~\ref{sec:ideal_ballooning_and_second_stabiity}), while the solid black line corresponds to the measured transition boundary $q^2 \beta_e = C_{\mathrm{nl}}^{\mathrm{STEP}}$ proposed in~\cite{zhang2026b}, where $C_{\mathrm{nl}}^{\mathrm{STEP}}$ is determined directly from a fine $\beta_e$ scan at fixed $q=3.5$, i.e.,\ from the location of the transition in $Q_{\mathrm{TOT}}$ (here near $(q, \beta_e) = (3.5, 0.026)$). The same threshold can alternatively be estimated from the quasilinear stress scaling of~\cite{zhang2026b} evaluated at a below-threshold reference simulation; this estimate agrees with the directly-measured value to within a factor of order unity but carries appreciable spread, since the quasilinear stress ratio departs from its scaling as the transition is approached. It will be shown in Section~\ref{subsec:forward_transition} (see also, Figure~\ref{fig:step_betascan}) that the increase in the heat flux is very sharp across the transition (i.e., there is a large jump in $Q_{\mathrm{TOT}}$ across a narrow window in $\beta_{e}$ at fixed $q$). The strong agreement between the predicted nonlinear threshold and the observed transition supports the relevance of the $q^2 \beta_e$ criterion in this regime \cite{zhang2026b}. The predicted nonlinear threshold provides a boundary for the local GK ``no-go zone'' (the region of red crosses), i.e., a region where local GK returns very large fluxes. The preferred STEP-EC-HD operating point is at $(q, \beta_e) = (3.5, 0.09)$ [the purple star in Figure~\ref{fig:q-beta-step}] which is in the EM high-transport region. We remind the reader that these simulations are performed with no equilibrium flow shear (NB STEP-EC-HD will have no external momentum injection); and that simulations with imposed flow shear can return much smaller fluxes~\cite{Giacomin2024,Giacomin_Dickinson_Dorland_Mandell_Bokshi_Casson_Dudding_Kennedy_Patel_Roach_2025}.

\begin{figure}
  \centering
\includegraphics[width=0.75\textwidth]{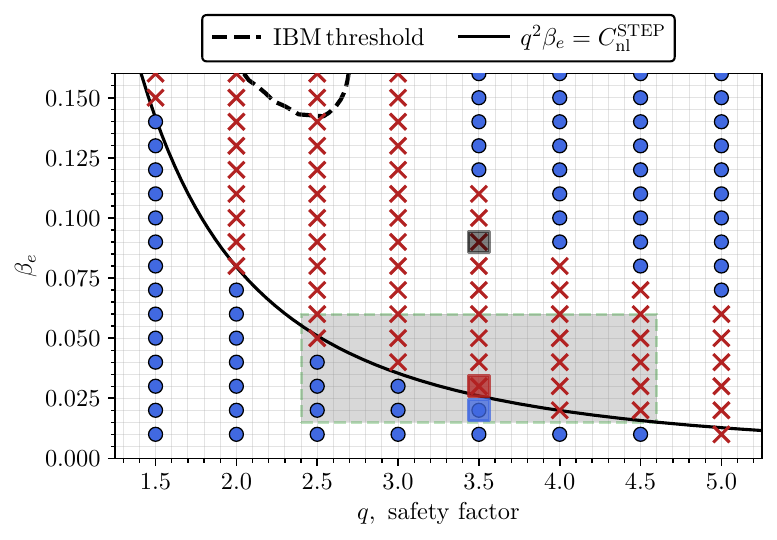}
  \caption{Results from nonlinear simulations of STEP-EC-HD at various values of $q$ and $\beta_{e}$ using \texttt{stella}. Other than $\beta^{\prime},$ which is always proportional to $\beta_{e},$ all other local parameters are fixed. Simulations which converge to $Q_{\mathrm{TOT}}/Q_{\mathrm{NZT}} < 1$ are marked by \textcolor{blue}{blue circles}. Simulations which do not converge (or which converge to fluxes $Q_{\mathrm{TOT}}/Q_{\mathrm{NZT}} > 1$) are marked by \textcolor{red}{red crosses}. The dashed black line (top left) is the ideal ballooning mode threshold calculated using \texttt{pyrokinetics}. The solid black line is the predicted transition curve $q^{2}\beta_{e} = C_{\mathrm{nl}}^{\mathrm{STEP}}$ \cite{zhang2026b} where $C_{\mathrm{nl}}^{\mathrm{STEP}}$ is computed by fitting a single simulation at $(q, \beta_{e}) = (3.5,0.026)$. The black box is the nominal STEP-EC-HD operating point. Blue and red boxes mark the locations of simulations analysed in Section~\ref{subsec:forward_transition}. A zoom of the shaded box is shown in Figure~\ref{fig:q-beta-step-transition-curve}.}
  \label{fig:q-beta-step}
\end{figure}

Interestingly, the scan in Figure~\ref{fig:q-beta-step} reveals \emph{two} sharp transitions that appear to bracket a region of large flux. This region is susceptible to transport bifurcations in the sense that, in the $\beta_{e}$ or $q$ directions, the flux is a non-monotonic function of $\beta_{e}$, so the same flux can be achieved at more than one value of $\beta_{e}$ (or $ \beta^{\prime}$). At lower values of $q^2 \beta_e$, simulations initially yield low heat fluxes before abruptly transitioning to high-transport states as $q$ or $\beta_e$ is increased. This marks the \emph{forward} transition identified in \cite{Pueschel2008,Pueschel2014} and studied in \cite{zhang2026b}. However, at slightly higher values of $q^2 \beta_e$, a \emph{reverse} transition is observed: at sufficiently large values of $q^{2}\beta_{e}$ the system can relax into a state of moderate turbulent transport. We note that this second transition is not specific to STEP and can be attributed to $\beta^\prime$ stabilisation~\cite{bourdelle2003}. The forward transition is discussed in Section~\ref{subsec:forward_transition} and the reverse transition is discussed in Section~\ref{subsec:reverse_transition}.

\subsection{Forward transition to extreme heat fluxes with increasing $\beta_{e}$; competition between \textbf{E}$\times$\textbf{B} stress and magnetic flutter stress defines a ``no-go zone'' for local gyrokinetics} \label{subsec:forward_transition}

\subsubsection{Below the non-zonal transition}
 Figures~\ref{fig:STEP_beta0p020_fig1}--\ref{fig:STEP_beta0p020_fig3} provide a detailed look at the nonlinear dynamics of STEP simulations just \emph{below} the predicted transition boundary, at $(q, \beta_{e}) = (3.5, 0.020)$ [the \textcolor{blue}{blue box} in Figure~\ref{fig:q-beta-step}]. In this low-$\beta_{e}$ regime, the turbulence saturates at relatively low heat flux levels, with the system exhibiting characteristics of an electrostatically dominated state. Figure~\ref{fig:STEP_beta0p020_fig1} shows the evolution of the electrostatic and electromagnetic electron and ion heat fluxes (left-hand side), and of the $k_{y}$ spectrum of the total heat flux (right-hand side). The simulation saturates at relatively modest fluxes, which are dominated by the electrostatic contribution. Figure~\ref{fig:STEP_beta0p020_fig2} shows the evolution of the $k_y$-spectrum of $\phi$ and $A_{\parallel}$, highlighting the persistence of a strong zonal component (black) and relatively much lower amplitude broadband activity in both $\phi$ and $A_{\parallel}$. This is indicative of a turbulence regime in which electromagnetic activity plays a subdominant role. Figure~\ref{fig:STEP_beta0p020_fig3} shows the time-averaged per-$k_{x}$ transfers (left-hand side) and net ZF torques (right-hand side) associated with the various stress terms defined in~(\ref{eq:lin_transfer})–(\ref{eq:bpar_transfer}). In this regime, the large-scale zonal flows are maintained by a balance between the $\bm{E} \times \bm{B}$ stress (zonal-flow reinforcing) and the sum of magnetic-flutter, compressive, and linear stresses (zonal-flow inhibiting). The result is a self-organised steady state in which zonal-flow-reinforcing and zonal-flow-inhibiting processes are delicately balanced, allowing a steady turbulent state in which large-amplitude zonal flows are able to saturate the turbulent fluxes at a moderate level. The time-averaged (over the shaded region in Figure~\ref{fig:STEP_beta0p020_fig2}, left-hand side) ZF torques are: $\langle T_\phi \rangle_{t} = 0.0139$, $\langle T_{A_\parallel}\rangle_{t}= -0.0024$, $\langle T_{\delta \! B_\parallel} \rangle_t = -0.0014$, and $\langle T_{\mathrm{lin}} \rangle_{t} = -0.0102$, yielding a total of O$(10^{-4})$. Below threshold, the electrostatic and linear physics dominate over any magnetic contributions to the torque. This dynamical equilibrium is disrupted as the transition boundary is approached.

In a statistically-steady saturated state, the zonal-flow energy budget closes with the stress injection balanced by damping, meaning that
\begin{equation}
\overline{\partial_t E_{\mathrm{ZF}}(k_x)} \;\simeq\; 0
\quad\Longrightarrow\quad
\overline{T_{\mathrm{lin},k_x}} \;+\; \overline{T_{\phi,k_x}} \;+\; \overline{T_{A_\parallel,k_x}} \;+\; \overline{T_{\delta B_\parallel,k_x}} \;\simeq\; 0,
\end{equation}
so the \emph{net transfer} into the ZF at each $k_x$ vanishes on average, consistent with saturated ZFs (stationary $\left|\langle\phi\rangle_{\psi,k_x}\right|$) where the positive nonlinear stresses balance the linear damping. In our simulations we observe in Figures~(\ref{fig:STEP_beta0p020_fig1})--(\ref{fig:STEP_beta0p020_fig3}) that a steady state exhibiting large ZFs relative to broadband drift-wave fluctuations necessarily follows an earlier growth interval during which the net transfer to ZFs was positive:
\begin{equation}
\Delta E_{\mathrm{ZF}}(k_x; t) \;=\; \int_{0}^{t} \mathrm{d}t^\prime \,
\big[T_{\mathrm{lin},k_x}(t^\prime)+T_{\phi,k_x}(t^\prime)+T_{A_\parallel,k_x}(t^\prime)+T_{\delta B_\parallel,k_x}(t^\prime)\big] \;>\; 0.
\end{equation}
That is, the nonlinear stresses must overcome the effective damping in $T_{\mathrm{lin},k_x}$ long enough to build up $E_{\mathrm{ZF}}(k_x)$ before time-averaged balance is achieved.

\begin{figure}
  \centering
  \begin{minipage}[t]{0.5\textwidth}
    \centering
    \includegraphics[]{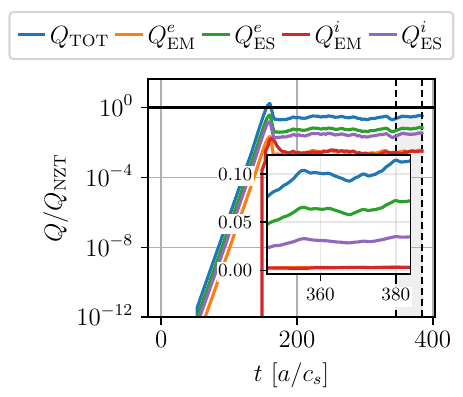}
  \end{minipage}\hfill
  \begin{minipage}[t]{0.5\textwidth}
    \centering
    \includegraphics[]{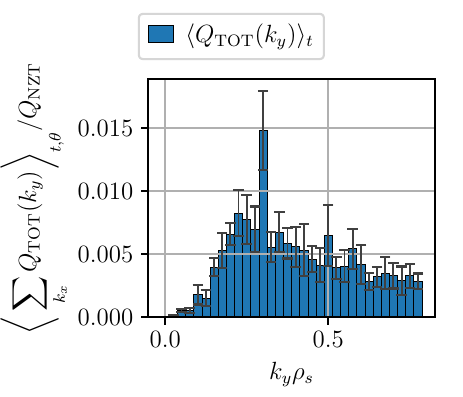}
\end{minipage}
  \caption{Results from nonlinear simulations of STEP at $(q, \beta_{e}) = (3.5,0.020),$ corresponding to the \textcolor{blue}{blue box} just \emph{below} the transition boundary in Figure~\ref{fig:q-beta-step}. Time traces of the electrostatic and electromagnetic electron and ion heat fluxes (left-hand side) and the time-averaged total heat-flux spectrum as a function of poloidal wavenumber $k_{y}$, normalised to $Q_{\mathrm{NZT}}$ and with error bars indicating the standard deviation over the averaging window (right-hand side). A zoom of the fluxes at late times is shown as an inset in the left-hand plot (note linear scale).
}
  \label{fig:STEP_beta0p020_fig1}
\end{figure}

\begin{figure}
  \centering
  \begin{minipage}[t]{0.5\textwidth}
    \centering
    \includegraphics[]{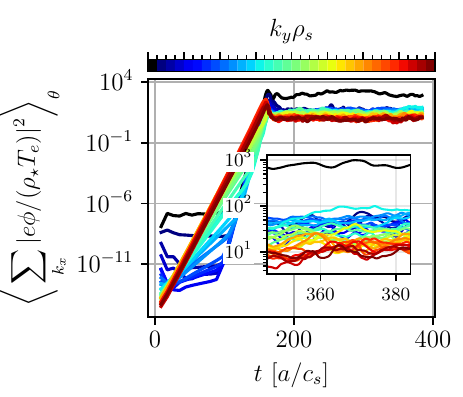}
  \end{minipage}\hfill
  \begin{minipage}[t]{0.5\textwidth}
    \centering
    \includegraphics[]{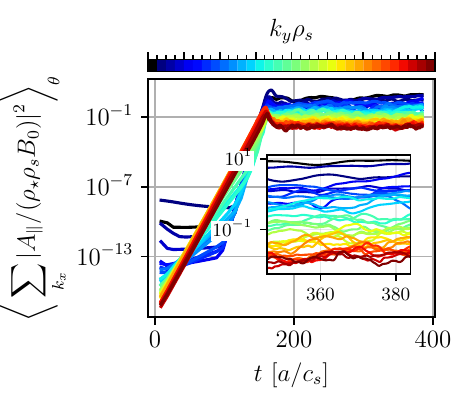}
\end{minipage}
  \caption{Results from nonlinear simulations of STEP at $(q, \beta_{e}) = (3.5,0.020),$ i.e., at the \textcolor{blue}{blue box} just \emph{below} the transition boundary in Figure~\ref{fig:q-beta-step}. Time traces of the $k_{y}$ spectrum of $\phi$ (left-hand side) and $A_{\parallel}$ (right-hand side). The zonal mode is shown in black. A zoom of the fields at late times is shown as an inset in each plot (note logarithmic scale).
}
  \label{fig:STEP_beta0p020_fig2}
\end{figure}

\begin{figure}
  \centering
  \begin{minipage}[t]{0.5\textwidth}
    \centering
    \includegraphics[]{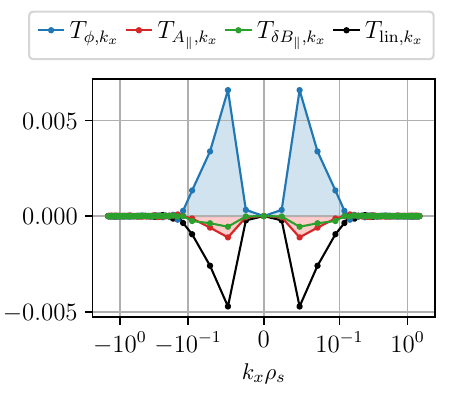}
  \end{minipage}\hfill
  \begin{minipage}[t]{0.5\textwidth}
    \centering
    \includegraphics[]{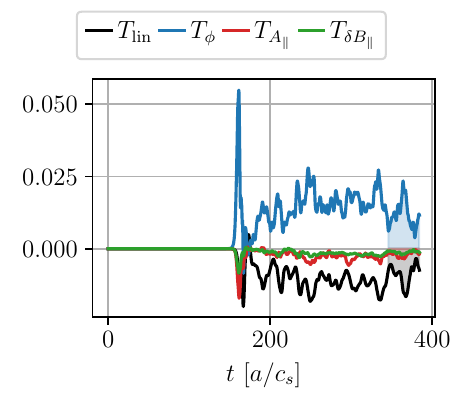}
  \end{minipage}
   \caption{Results from nonlinear simulations of STEP at $(q, \beta_{e}) = (3.5,0.020),$ corresponding to the \textcolor{blue}{blue box} just \emph{below} the transition boundary in Figure~\ref{fig:q-beta-step}. Time-averaged (over the grey shaded region in Figure~\ref{fig:STEP_beta0p020_fig1}, left-hand side) spectrum of the zonal transfers where shading denotes the large-scale cutoff used in (\ref{eq:cutoff}) (left-hand side). Time traces of the large-scale zonal torques defined in (\ref{eq:lin_transfer})--(\ref{eq:apar_transfer}) (right-hand side).
}
  \label{fig:STEP_beta0p020_fig3}
\end{figure}

\begin{figure}
  \centering
  \begin{minipage}[t]{0.5\textwidth}
    \centering
    \includegraphics[]{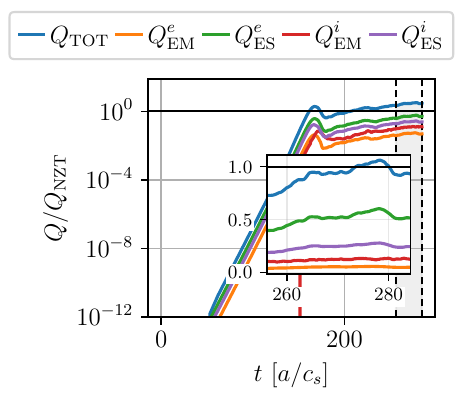}
  \end{minipage}\hfill
  \begin{minipage}[t]{0.5\textwidth}
    \centering
    \includegraphics[]{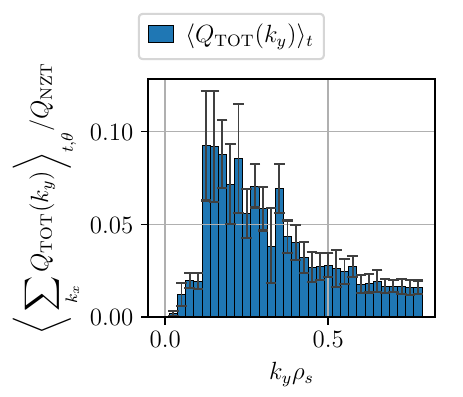}
\end{minipage}
  \caption{Results from nonlinear simulations of STEP at $(q, \beta_{e}) = (3.5,0.030),$ corresponding to the \textcolor{red}{red box} just \emph{above} the transition boundary in Figure~\ref{fig:q-beta-step}. Time traces of the electrostatic and electromagnetic electron and ion heat fluxes (left-hand side) and the time-averaged total heat-flux spectrum as a function of poloidal wavenumber $k_{y}$, normalised to $Q_{\mathrm{NZT}}$ and with error bars indicating the standard deviation over the averaging window (right-hand side). The heat flux is increasing slowly even at late times in the simulation, and the time average is taken over the grey shaded region. A zoom of the fluxes at late times is shown as an inset in the left-hand plot (note linear scale).
}
  \label{fig:STEP_beta0p030_fig1}
\end{figure}

\begin{figure}
  \centering
  \begin{minipage}[t]{0.5\textwidth}
    \centering
    \includegraphics[]{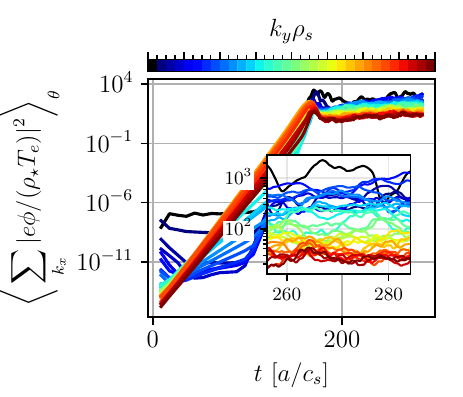}
  \end{minipage}\hfill
  \begin{minipage}[t]{0.5\textwidth}
    \centering
    \includegraphics[]{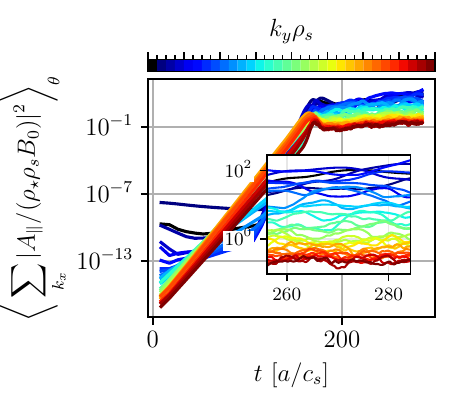}
\end{minipage}
  \caption{Results from nonlinear simulations of STEP at $(q, \beta_{e}) = (3.5,0.030),$ corresponding to the \textcolor{red}{red box} just \emph{above} the transition boundary in Figure~\ref{fig:q-beta-step}. Time traces of the $k_{y}$ spectrum of $\phi$ (left-hand side) and $A_{\parallel}$ (right-hand side). The zonal mode is shown in black.  A zoom of the fields over the time-averaged region is shown as an inset in each plot (note logarithmic scale).
}

  \label{fig:STEP_beta0p030_fig2}
\end{figure}

\begin{figure}
  \centering
  \begin{minipage}[t]{0.49\textwidth}
    \centering
    \includegraphics[width=\textwidth]{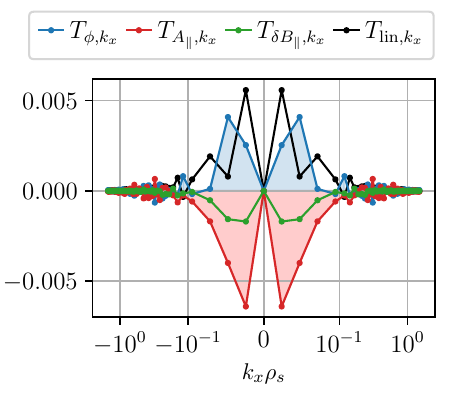}
  \end{minipage}\hfill
  \begin{minipage}[t]{0.49\textwidth}
    \centering
    \includegraphics[width=\textwidth]{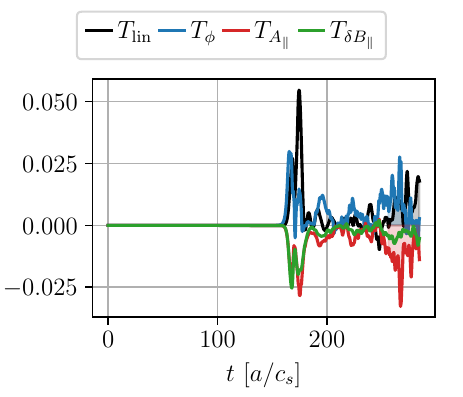}
  \end{minipage}
   \caption{Results from nonlinear simulations of STEP at $(q, \beta_{e}) = (3.5,0.030),$ corresponding to the \textcolor{red}{red box} just \emph{above} the transition boundary in Figure~\ref{fig:q-beta-step}. Time-averaged (over the grey shaded region in Figure~\ref{fig:STEP_beta0p030_fig1}, left-hand side) $k_{x}$ spectrum of the zonal transfers where shading denotes the large-scale cutoff used in (\ref{eq:cutoff}) (left-hand side). Time traces of the large-scale zonal transfers defined in (\ref{eq:lin_transfer})--(\ref{eq:apar_transfer}) (right-hand side).
}
  \label{fig:STEP_beta0p030_fig3}
\end{figure}

\subsubsection{Above the non-zonal transition}
Figures~\ref{fig:STEP_beta0p030_fig1}--\ref{fig:STEP_beta0p030_fig3} show STEP simulations just \emph{above} the predicted transition boundary, at $(q, \beta_{e}) = (3.5, 0.030)$ [the \textcolor{red}{red box} in Figure~\ref{fig:q-beta-step}]. Above the transition threshold, the non-zonal turbulent fluctuations reach higher amplitudes than they do below threshold and are less dominated by ZFs.  The turbulent fluxes in Figure~\ref{fig:STEP_beta0p030_fig1} are an order of magnitude larger and continue to increase slowly even at late times in the simulation. Figure~\ref{fig:STEP_beta0p030_fig2} shows the evolution of the $k_y$-spectrum of $\phi$ and $A_{\parallel}$, where it can be seen that the non-zonal components of $\phi$ continue to grow to late times in the simulation and, unlike below threshold, even overtake the zonal component towards the end of the simulation period (see zoomed inset). This is the key signature of the non-zonal transition~\cite{Pueschel2013}. Figure~\ref{fig:STEP_beta0p030_fig3} shows the time-averaged per-$k_{x}$ transfers (left-hand side) and net ZF torques (right-hand side) associated with the various stress terms defined in~(\ref{eq:lin_transfer})–(\ref{eq:bpar_transfer}). Above the transition, the magnetic-flutter stress (zonal flow inhibiting) exceeds the $\bm{E} \times \bm{B}$ stress (zonal flow reinforcing) and the steady state is lost.

\begin{figure}
    \centering
    \includegraphics[width=0.75\linewidth]{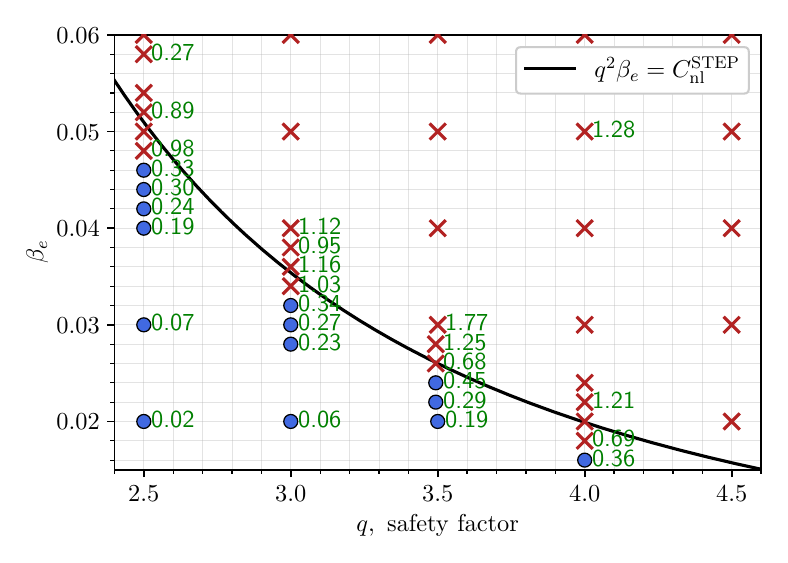}
    \caption{Zoom of the shaded box in Figure~\ref{fig:q-beta-step} focussing on the region around the predicted transition curve. The \textcolor{ForestGreen}{green numbers} denote the value of the time-averaged ratio $|\langle 
T_{A_\parallel}\rangle_{t}/\langle T_\phi\rangle_{t}|$. The boundary predicted by \cite{zhang2026b} fits the simulation results remarkably well.}
    \label{fig:q-beta-step-transition-curve}
\end{figure}

\subsubsection{Threshold for the non-zonal transition}
Figure~\ref{fig:q-beta-step-transition-curve} provides a magnified view of Figure~\ref{fig:q-beta-step} (a zoom of the grey box), focusing on the region near the predicted nonlinear transition boundary $q^2 \beta_e = C_{\mathrm{nl}}$. This zoomed perspective highlights the transition between simulations that converge to low-flux states and those that do not. As in Figure~\ref{fig:q-beta-step}, the markers denote whether each simulation converged within the simulated time window, and the green annotations indicate the time-averaged ratio $|\langle T_{A_\parallel} \rangle_t / \langle T_\phi \rangle_t|$. The numerator and denominator in this ratio \emph{always} have the opposite sign in these simulations. Wherever they are reported, time-averages are taken over the last 10\% of the simulations (as earlier, simulations are shown with a blue marker if the heat flux is statistically stationary in this window and below the chosen threshold and with a red marker if either of these conditions are not satisfied). The diagnostic $|\langle T_{A_\parallel} \rangle_t / \langle T_\phi \rangle_t|$ serves as a proxy for the relative contribution of electromagnetic versus electrostatic zonal stress terms in the saturated state and is predicted to pass through unity at the transition threshold \cite{zhang2026b}. The solid black line is a fit of $q^{2}\beta_{e} = C_{\mathrm{nl}}$ through a single point at $(q,\beta_{e}) = (3.5, 0.026).$ It can be seen that the $q^2 \beta_e$ scaling proposed in~\cite{zhang2026b} fits the simulation results remarkably well, underscoring the predictive utility of this scaling. 

\subsubsection{On the role of linear instabilities in the forward transition}

We confirm the findings of many authors \cite{zhang2026b, Pueschel2008,Pueschel2014} that there is no obvious change in the linear physics across the transition threshold. Figure~\ref{fig:STEPlinear} shows the variation of growth rate (left-hand side) and mode frequency (right-hand side) with \( k_y \rho_s \), of the fastest growing linear mode based on initial value simulations of STEP-EC-HD at several values of \( \beta_e \), all performed at fixed safety factor \( q = 3.5 \). The chosen \( \beta_e \) values correspond to those used in the nonlinear simulations of Figure~\ref{fig:q-beta-step-transition-curve}, crossing the transition boundary. Across the full range of \( k_y \rho_s \), which spans toroidal mode numbers from \( n = 1 \) to \( n = 300 \), the dominant linear modes exhibit no significant qualitative change in either growth rate or real frequency as \( \beta_e \) increases. This separates the forward transition from the reverse transition where a change in microstability is observed (see Section~\ref{subsubsec:linear_reverse_transition}).

In an attempt to identify subdominant growing modes, we applied \emph{Dynamic Mode Decomposition} (DMD)~\cite{Schmid2010} to the time-evolving fields from the same set of linear gyrokinetic simulations. DMD is a data-driven technique that approximates the spectral content of a system by decomposing time-series data into modes with well-defined complex frequencies and growth rates. We were able to find no unstable subdominant linear modes (note that the threshold is at much lower $\beta_{e}$ than the nominal operating point where $\beta_e \sim 0.09$ and unstable subdominant MTMs were reported~\cite{kennedy2023a})\footnote{It is worth remarking that applying DMD to these simulations can be challenging and it is possible for DMD to fail if the dominant instability recovered by the initial value solver dominates the fields. In such cases, all snapshots become scalar multiples of the same structure, meaning that the data matrix becomes effectively rank-1, and hence ill-conditioned.}. The DMD results were spot checked using the eigenvalue solver in \texttt{GENE}, which also found no unstable subdominant modes for these parameter sets. We regard this as indicative rather than definitive: DMD can become ill-conditioned when a single instability dominates the signal (see footnote), so a fully converged enumeration of subdominant modes near threshold would require a dedicated eigenvalue study; the corroborating \texttt{GENE} eigenvalue calculations were performed at representative $(k_y,\beta_e)$ spanning the transition.  

\begin{figure}
  \centering
  \begin{minipage}[t]{0.49\textwidth}
    \centering
    \includegraphics[width=\textwidth]{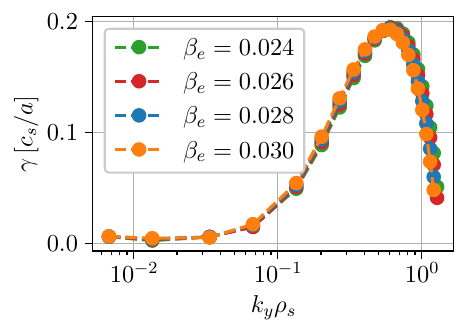}
  \end{minipage}\hfill
  \begin{minipage}[t]{0.49\textwidth}
    \centering
    \includegraphics[width=\textwidth]{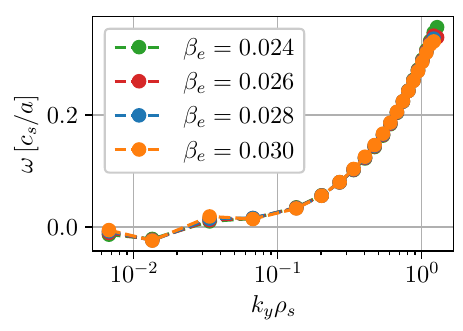}
  \end{minipage}
   \caption{Growth rate (left-hand side) and mode frequency (right-hand side) as functions of $k_y \rho_s$ from \texttt{stella} linear simulations of STEP-EC-HD at various values of $\beta_{e}$ for fixed $q = 3.5$. The values of $\beta_{e}$ chosen correspond to the values at which the nonlinear transition is observed in Figure~\ref{fig:q-beta-step-transition-curve}. The considered $k_y$ values cover a range corresponding to toroidal mode numbers between $n=1$ and $n=300$.  Solid and open markers refer to unstable and stable modes, respectively. {The growth rate of stable modes is set to zero.}
}
\label{fig:STEPlinear}
\end{figure}

\subsubsection{Forward transition as a constraint on $\beta_{\mathrm{pol}}$}
\label{subsubsec:forward_transition_as_a_constraint_on_betapol}

As discussed in Section~\ref{sec:tokamak_threshold}, the condition on $q^2\beta_e$ can be thought of as a condition on $\beta_{\mathrm{pol}}.$ We verify this by varying the major radius $R_{\mathrm{maj}}$ at fixed minor radius $a$ and fixed $r/a,$ which corresponds to moving the flux surface to larger major radius with all other local parameters held fixed. Figure~\ref{fig:Rmajor_scan} (left-hand side) demonstrates that increasing the major radius systematically reduces the critical $\beta_{e}$ required for the onset of the non-zonal transition, indicating that higher aspect ratio plasmas are more prone to entering the non-zonal regime at lower pressure (as observed in \cite{zhang2026b}). At fixed $(q,\beta_{e})=(3.5,0.024)$, the dominant positive ZF torque $T_{\phi}$ is progressively counteracted by increasingly strong zonal-flow-depleting contributions from $T_{A_\parallel}$ and $T_{\delta \! B_\parallel}$ with $R_{\mathrm{maj}}$ (Figure~\ref{fig:Rmajor_scan}, right-hand side).  This result suggests that this physics becomes increasingly important as the aspect ratio increases.

Crucially, this $R_{\mathrm{maj}}$ scan discriminates a genuine $\beta_{\mathrm{pol}}$ constraint from a mere dependence on aspect ratio. A criterion in $q^2\beta_e$ alone, with $C_{\mathrm{nl}}$ held fixed, would give $\beta_{e,\mathrm{crit}} = C_{\mathrm{nl}}/q^2$ \emph{independent} of $R$ at fixed $q.$ Instead, $\beta_{e,\mathrm{crit}}$ is observed to \emph{fall} with $R,$ with the magnetic-flutter-to-$\bm{E}\times\bm{B}$ torque ratio $|T_{A_\parallel}|/|T_\phi|$ rising correspondingly; by~(\ref{eq:q2beta_betapol_main}) this decrease is precisely the additional factor of $\epsilon^2$ supplied by the poloidal-beta limit $\beta_{\mathrm{pol,crit}}\sim C_{\mathrm{nl}}/\epsilon^2.$ The controlling parameter is thus $\beta_{\mathrm{pol}}$ (equivalently $q^2\beta_e$), and the $R$-dependence is the signature of that constraint rather than an independent aspect-ratio effect.

\begin{figure}
    \centering
    \includegraphics[width=\linewidth]{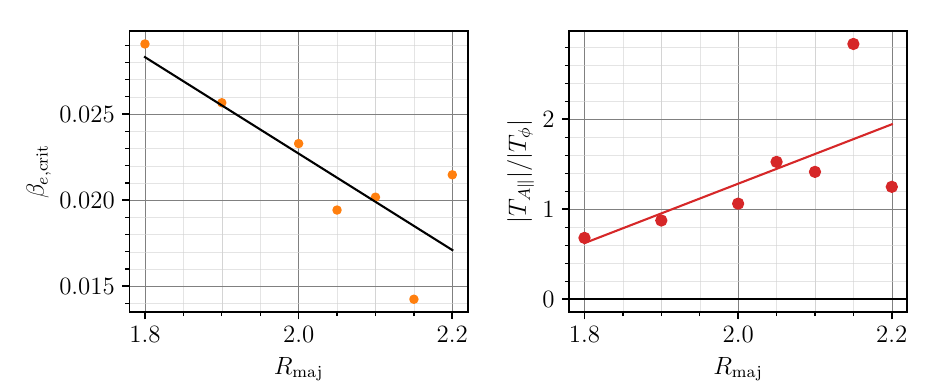}
        
    \caption{Critical electron beta, $\beta_{e,\mathrm{crit}}$, for the onset of the non-zonal state as a function of major radius $R_{\mathrm{maj}}$, inferred from a single low-$\beta_e$ nonlinear simulation at fixed $(q,\beta_e)=(3.5,0.024)$ for each $R$ using the ZF torque scaling of~\cite{zhang2026b} (left-hand side). Magnitude of the time-averaged ZF torques measured from nonlinear simulations at fixed $(q,\beta_e)=(3.5,0.024)$ and plotted as a function of $R_{\mathrm{maj}}$ (right-hand side). A least squares fit is shown for each plot.}
    \label{fig:Rmajor_scan}
\end{figure}

\subsection{Reverse transition to lower heat fluxes with increasing $\beta_{e}$; the role of $\beta^{\prime}$ stabilisation}
\label{subsec:reverse_transition}

Given the generality of the arguments presented in \cite{zhang2026a,zhang2026b}, it may be surprising to find the existence of a lower-flux state at larger values of $q^{2}\beta_{e}$ in Figure~\ref{fig:q-beta-step} (blue circles in the upper-right-hand corner). In \cite{zhang2026b} (and in Section~\ref{subsec:forward_transition}), the magnetic-flutter stress was shown to dominate the zonal-flow dynamics in the high-$\beta_e$ regime, typically leading to strong suppression of zonal flows and correspondingly large transport levels once a critical value of $q^2\beta_{e}$ was exceeded. However, the situation is more subtle when we allow access to even larger pressure gradients. 

We first explore (and verify) this phenomenon numerically before giving an intuitive explanation in Section~\ref{subsubsec:linear_reverse_transition}.

\subsubsection{Reverse Transition in CBC}

Our first test of this theory is to conduct nonlinear simulations of CBC geometry \cite{Dimits2000}, identical to those performed in \cite{zhang2026b} but at larger values of $\beta_{e}.$  Figures~\ref{fig:CBC_highbeta_secondstability_fig1} and \ref{fig:CBC_highbeta_secondstability_fig2} show results from nonlinear simulations of CBC geometry at \( (q, \beta_{e}) = (2, 0.04) \). This parameter point lies well above the nonlinear threshold but remains IBM-stable. In Figure~\ref{fig:CBC_highbeta_secondstability_fig1}, time traces of the \( k_{y} \) spectrum of the electrostatic potential \( \phi \) and the parallel vector potential \( A_{\parallel} \) reveal a broadband turbulent state that includes significant low-\( k_y \) activity and strong zonal modes (black lines). The time traces are steady over $1000$ ion sound times and reach a robustly steady saturated state at reasonable fluxes [i.e., $O(10^2)\,\mathrm{kW\,m^{-2}}$ for DIII-D parameters, corresponding to roughly $2\,\mathrm{MW}$ of power crossing the flux surface (taking $R_0\approx1.67\,\mathrm{m}$ and $r/a=0.5$, a flux-surface area of $\approx20\,\mathrm{m^2}$), a transport level comparable to standard Cyclone Base Case ITG studies~\cite{Dimits2000}].  Figure~\ref{fig:CBC_highbeta_secondstability_fig2} shows that in this high-$\beta_{e}$ saturated regime, the turbulent heat flux remains steady. The right-hand side illustrates that the magnetic-flutter stress, which damped zonal flows at lower $\beta$ below and just above threshold, is now able to reinforce ZFs transiently. In the saturated phase the stress injection is balanced by damping. When compared to the lower-$\beta_e$ electromagnetic high-transport state simulations in \cite{zhang2026b}, the stresses are able to change sign (in particular the linear stress can become zonal flow reinforcing) and different dynamic balances can be sustained.

\begin{figure}
  \centering
  \begin{minipage}[t]{0.5\textwidth}
    \centering
    \includegraphics[]{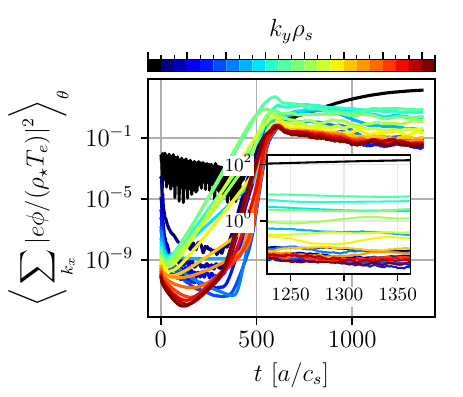}
  \end{minipage}\hfill
  \begin{minipage}[t]{0.5\textwidth}
    \centering
    \includegraphics[]{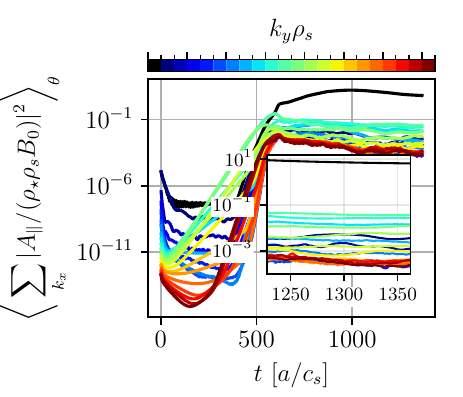}
\end{minipage}
  \caption{Results from nonlinear simulations of CBC geometry at $(q, \beta_{e}) = (2,0.04)$. Time traces of the $k_{y}$ spectrum of $\phi$ (left-hand side) and $A_{\parallel}$ (right-hand side). The zonal mode is shown in black. A zoom of the fields at late times is shown as an inset in each plot (note logarithmic scale).
}
  \label{fig:CBC_highbeta_secondstability_fig1}
\end{figure}

\begin{figure}
  \centering
  \begin{minipage}[t]{0.5\textwidth}
    \centering
    \includegraphics[]{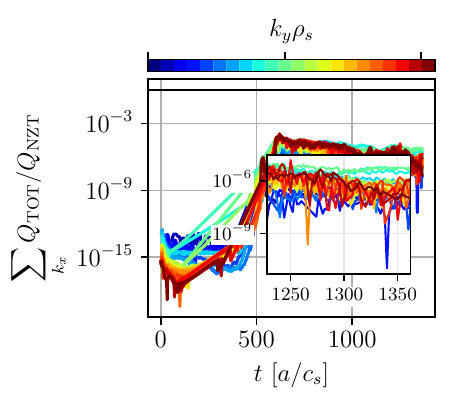}
  \end{minipage}\hfill
  \begin{minipage}[t]{0.5\textwidth}
    \centering
    \includegraphics[]{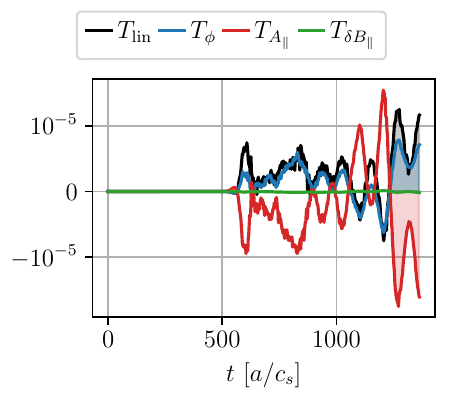}
  \end{minipage}
   \caption{Results from nonlinear simulations of CBC geometry at $(q, \beta_{e}) = (2,0.04)$. Time traces of the total heat flux partitioned by poloidal wavenumber (left-hand side) and the large-scale ZF torques defined in (\ref{eq:lin_transfer})-(\ref{eq:apar_transfer}) (right-hand side).   
}
  \label{fig:CBC_highbeta_secondstability_fig2}
\end{figure}

\subsubsection{Reverse Transition in STEP-EC-HD}

Figure~\ref{fig:q-beta-step} suggests the existence of a high-$q^{2}\beta_{e}$ region for STEP-EC-HD where simulations can saturate at reasonable fluxes. Figure~\ref{fig:q-beta-step-transition-partitioned} shows the (signed) time-integrated (over the entire simulation) zonal flow torques as a function of \( q^2\beta_e \) for a subset\footnote{Runs in which the turbulence is fully extinguished ($Q<0.1~\mathrm{MW\,m^{-2}}$) are excluded from Figure~\ref{fig:q-beta-step-transition-partitioned}, since there the zonal stresses are numerical noise and the (residual) linear torque merely tracks a slow drift of the surviving zonal field rather than a saturated balance.} of the simulations presented in Figure~\ref{fig:q-beta-step}, with the vertical dashed line indicating the critical value \( q^2\beta_e = C_{\mathrm{nl}}^{\mathrm{STEP}} \), above which a transition to non-zonal, high-transport states occurs. The plot shows three physically distinct regimes: the pre-transition regime, where turbulence is zonal-dominated and heat fluxes are relatively low [panels (a), (d), and (g)]; the electromagnetic transition phase, marked by a sharp rise in heat flux and breakdown of zonal regulation [panels (b), (e), and (h)]; and the post-transition regime, where partial restoration of zonal structure is observed [panels (c), (f), and (i)].

In panel (a) of Figure~\ref{fig:q-beta-step-transition-partitioned} (top row),  the electromagnetic contributions (green triangles and red triangles) to the stress are small, and the dominant balance is between the $\bm{E} \times \bm{B}$ stress (blue circles) that reinforces zonal flows, and the linear stress (black squares) that damps them. For small \( q^2\beta_e \), these components nearly cancel, yielding total transfers (black crosses) that are slightly positive [panel (d)]. The fluxes are modest in this zonally regulated regime [panel (g)]. As \( q^2\beta_e \) increases, both the magnetic-flutter stress (red triangles) and compressive stress (green triangles) oppose the primary electrostatic $\bm{E} \times \bm{B}$ stress, and these grow in magnitude. Near the transition point [far right of panel (a), far left of panel (b)], the total transfer becomes net negative [far right of panel (d), far left of panel (e)], indicating the suppression of zonal flows and weakening their effectiveness to saturate the turbulence. This is accompanied by the transition to a large flux state [far right of panel (h), far left of panel (i)]. At even larger values of $q^{2}\beta_e$ all of the stresses get very small [panel (c)] and the total transfer once again becomes close to zero [panel (f)] allowing saturation, which occurs at moderate fluxes [panel (i)].

\begin{figure}
    \centering
    \includegraphics[]{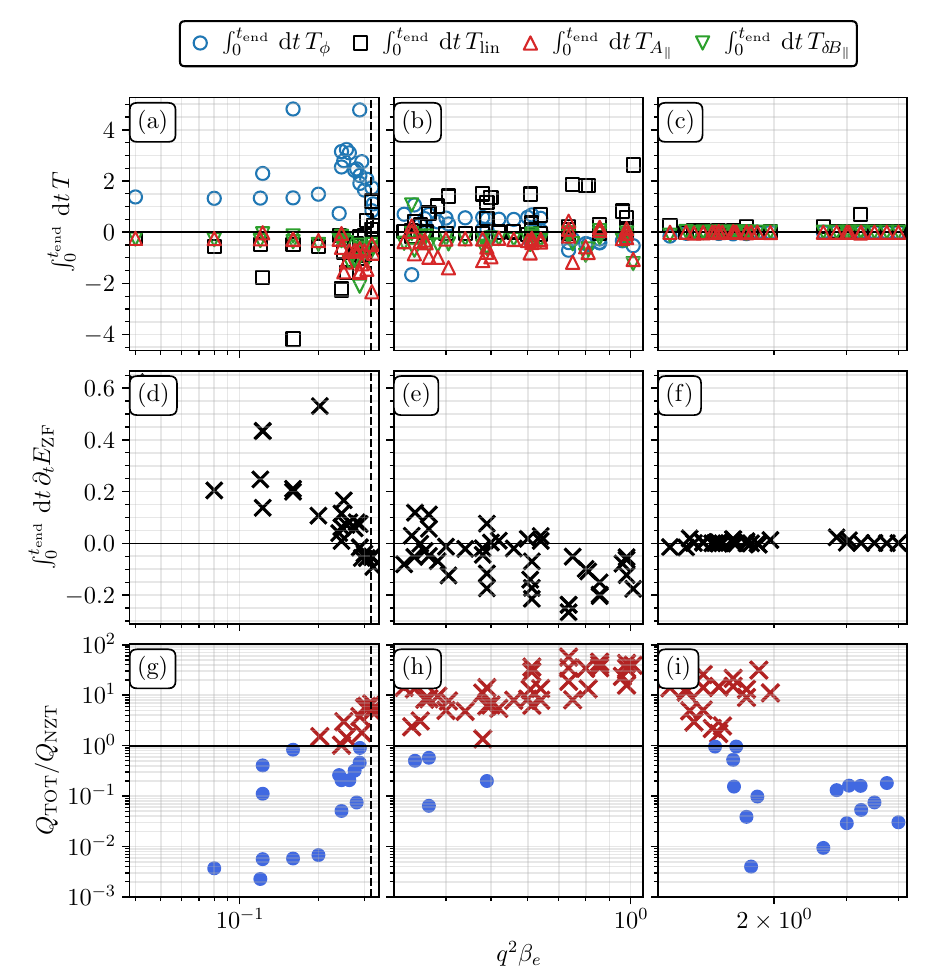}
    \caption{Time integrated (over the entire simulation) contributions to the zonal flow torque (top row) and total zonal flow torque (middle row). Time averaged (over the final 10\% of each simulation) normalised heat flux (bottom row) as a function of $q^{2}\beta_{e}$ for the data points in Figure~\ref{fig:q-beta-step}. The black dashed vertical line is $q^{2}\beta_{e} = C_{\mathrm{nl}}^{\mathrm{STEP}}.$ Each column corresponds to a physically distinct regime: before the non-zonal transition [(a), (d), and (g)]; during the electromagnetic transition [(b), (e), and (h)]; and after the reverse transition [(c), (f), and (i)]. In the bottom row, simulations which converge to $Q_{\mathrm{TOT}}/Q_{\mathrm{NZT}} < 1$ are marked by \textcolor{blue}{blue circles}. Simulations which do not converge (or which converge to fluxes $Q_{\mathrm{TOT}}/Q_{\mathrm{NZT}} > 1$) are marked by \textcolor{red}{red crosses}.}
    \label{fig:q-beta-step-transition-partitioned}
\end{figure}

\subsubsection{On the role of linear instabilities and $\beta^\prime$ stabilisation in the reverse transition}
\label{subsubsec:linear_reverse_transition}

The reverse transition and the weakening of the stresses identified in Figure~\ref{fig:q-beta-step-transition-partitioned} are due to the linear physics of the system. 

Figure~\ref{fig:STEPlinear_reversetransition} shows the linear growth rates (left-hand side) and real frequencies (right-hand side) of the dominant instability for a range of $k_y \rho_s$ values as functions of $\beta_e$, obtained from \texttt{GENE} simulations of the STEP-EC-HD equilibrium at fixed $q = 3.5$. These simulations span the region of $\beta_e$ associated with the reverse transition observed in Figure~\ref{fig:q-beta-step}. In the left panel, we observe that at low $\beta_e$, hKBMs dominate the linear spectrum, characterised by relatively large growth rates and positive mode frequencies (right panel). As $\beta_e$ increases, the hKBM growth rates are suppressed by $\beta^{\prime}$ stabilisation~\cite{bourdelle2005}, and a new instability branch emerges; this mode is the MTM that was reported to be subdominant at the nominal parameters~\cite{kennedy2023a}. 

\begin{figure}
  \centering
  \begin{minipage}[t]{0.5\textwidth}
    \centering
    \includegraphics[]{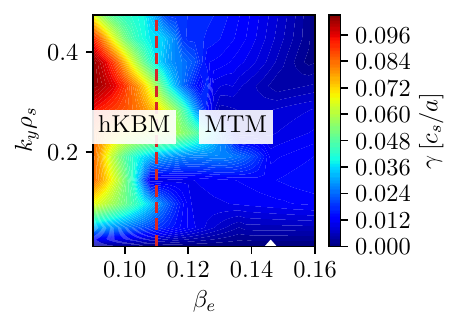}
  \end{minipage}\hfill
  \begin{minipage}[t]{0.5\textwidth}
    \centering
    \includegraphics[]{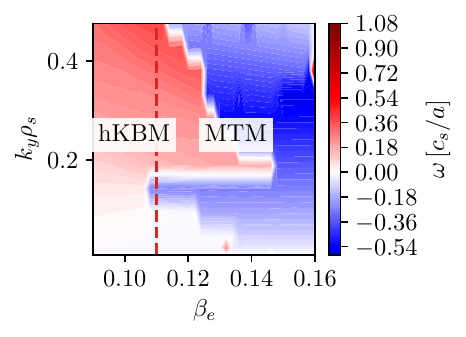}
  \end{minipage}
   \caption{Growth rate (left-hand side) and mode frequency (right-hand side) as functions of $k_y \rho_s$ and $\beta_{e}$ from \texttt{GENE} linear simulations of STEP-EC-HD for fixed $q = 3.5$. The values of $\beta_{e}$ cover the reverse transition observed in Figure~\ref{fig:q-beta-step}. The red line denotes the value of $\beta_{e}$ for which $Q_{\mathrm{TOT}} < Q_{\mathrm{NZT}}$ in Figure~\ref{fig:step_betascan}. The considered $k_y$ values cover a range corresponding to toroidal mode numbers between $n=1$ and $n=300$. {The growth rate of stable modes is set to zero.}
}
\label{fig:STEPlinear_reversetransition}
\end{figure}

To more easily see this change in the nature of the dominant instability we introduce the linear tearing parameter, $C_{\mathrm{tear}},$ defined as the ratio
\begin{equation}
C_{\mathrm{tear}} = \frac{\left|\, \displaystyle\int \mathrm{d}\theta\, A_\parallel(\theta)\, \sqrt{g_{\theta\theta}} \,\right|}{\displaystyle\int \mathrm{d}\theta\, \bigl|A_\parallel(\theta)\bigr|\, \sqrt{g_{\theta\theta}}},
\label{eq:Ctear}
\end{equation}
where $A_\parallel(\theta)$ is the real part of the parallel vector potential along the field line and $g_{\theta\theta} \equiv \nabla\theta\cdot\nabla\theta$ is the field-aligned covariant metric component relating poloidal angle increments to arc length. This parameter quantifies whether a perturbation results in a field line that does not
return to the equilibrium flux surface: values near zero correspond to antisymmetric (odd) parity, while values near unity indicate symmetric (even) parity (e.g., \cite{patel2021}). Figure~\ref{fig:ctear_reverse_transition} shows the value of the linear tearing parameter for the linear simulations in Figure~\ref{fig:STEPlinear_reversetransition}. It can be seen that the dominant mode transitions from odd parity to even parity, i.e., from twisting modes to tearing modes, as $\beta_{e}$ is increased.  

\begin{figure}
    \centering
    \includegraphics[]{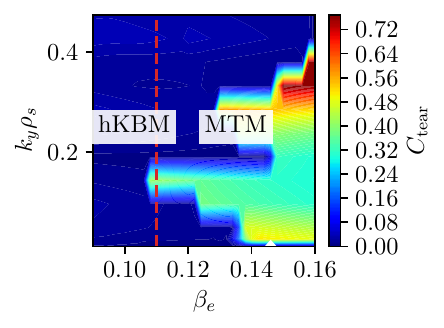}
    \caption{Value of the linear tearing parameter $C_{\mathrm{tear}}$ (\ref{eq:Ctear}) as a function of $k_{y}\rho_{s}$ and $\beta_{e}$ from \texttt{GENE} linear simulations of STEP-EC-HD for fixed $q = 3.5$. The red line denotes the value of $\beta_{e}$ for which $Q_{\mathrm{TOT}} < Q_{\mathrm{NZT}}$ in Figure~\ref{fig:step_betascan}.}
    \label{fig:ctear_reverse_transition}
\end{figure}

This linear stabilisation has a direct counterpart in the zonal-flow stress balance, and its prominence here appears to be geometric in origin. In the reverse-transition regime the turbulent stresses are quenched together with the hKBM drive [panels~(c) and~(f) of Figure~\ref{fig:q-beta-step-transition-partitioned}], so the saturated zonal-flow budget is closed instead by the linear stress $\Pi_{\mathrm{lin}},$ which has become zonal-flow-reinforcing. Measured directly from the $q=3.5$ scan, the net linear torque $T_{\mathrm{lin}}=\mathrm{Re}\,\langle\phi\rangle^{*}\Pi_{\mathrm{lin}}$ (summed over the large-scale zonal modes) reverses sign across the forward transition, from zonal-flow-damping below and at threshold [$T_{\mathrm{lin}}\approx-1.1\times10^{-2}$ at $q^{2}\beta_{e}=0.25$] to zonal-flow-reinforcing beyond it [$+7.7\times10^{-3}$ at $q^{2}\beta_{e}=0.61$ and $+1.2\times10^{-3}$ at $q^{2}\beta_{e}=1.23$]. As shown in~\ref{app:linear_term}, the curvature-driven part of $\Pi_{\mathrm{lin}}$ is controlled by the geometric ratio $B_{\mathrm{pol}}/B_{\mathrm{tor}}\sim\epsilon/q\sim r/qR$ (here $B_{\mathrm{pol}}$ is the poloidal field $B_\theta$): it is negligible when this ratio is small (large aspect ratio and/or large $q$) and becomes significant, and sensitive to flux-surface shaping, when it is not. The reverse transition is influenced by the same flux-surface shaping that controls ideal-ballooning stability (Section~\ref{sec:ideal_ballooning_and_second_stabiity}), since this shaping also sets the size of the linear stress that ultimately closes the zonal-flow budget. This is consistent with the moderate-flux, second-stable state being reached in the tight-aspect-ratio geometry of STEP-EC-HD, and suggests that its accessibility might be influenced through shaping.

\subsection{Comparing the forward and reverse transitions}

Table~\ref{tab:forward_reverse_mechanisms} highlights the contrasting physics of the forward and reverse transitions. The forward transition is a genuinely nonlinear effect, effectively an EM Dimits shift, in which the balance of stresses identified in \cite{zhang2026a,zhang2026b} breaks down once $q^{2}\beta_{e}$ exceeds a critical value. Because the threshold is set by a subtle reorganisation of nonlinear stress transfers, quasilinear (QL) models may fail to capture its onset reliably\footnote{Note that it is more important for reduced transport models to capture the location of the threshold than to predict with accuracy the fluxes well above, since these are likely to be much higher than the available heating power and therefore inaccessible.}. In contrast, the reverse transition at larger $q^{2}\beta_{e}$ is fundamentally linear in nature even though its signature can be seen in the stress balance. The reverse transition is driven by the $\beta^{\prime}$ stabilisation mechanism~\cite{bourdelle2003}, which suppresses the hKBM and allows microtearing-like modes to become dominant. This distinction is important for predictive modelling. If future devices such as STEP are to operate at large $q^{2}\beta_{e}$ without relying on strong equilibrium flow shear, then ensuring access to the reverse-transition regime is crucial to avoid large electromagnetic transport. The fact that this transition is underpinned by a linear stabilisation mechanism means that it should be within reach of QL models, providing an efficient route for integrated modelling to identify and optimise operating points in this high-$\beta$ regime. In the next section we connect the reverse transition to the theory of the ideal ballooning mode - potentially contributing to a fast proxy for use in integrated modelling. 

\begin{table}
    \centering
    \renewcommand{\arraystretch}{1.2}
    \begin{tabular}{p{3.6cm} p{5.2cm} p{5.2cm}}
        \toprule
        & \textbf{Forward transition} & \textbf{Reverse transition} \\
        \midrule
        \textbf{Mechanism} &
        Nonlinear  &
        Linear \\ 
        
        \textbf{Explanation} &
        Stress-balance mechanism of \cite{zhang2026a,zhang2026b} &
        $\beta^{\prime}$ stabilisation of \cite{bourdelle2003} \\ 
        
        \textbf{QL models} &
        May fail to capture the threshold &
        Can capture linear physics \\
        \bottomrule
    \end{tabular}
    \caption{Comparison of forward and reverse transitions in electromagnetic turbulence.}
    \label{tab:forward_reverse_mechanisms}
\end{table}

\section{Relationship to ideal ballooning and access to second stability}
\label{sec:ideal_ballooning_and_second_stabiity}

In Section~\ref{sec:numerical_simulations}, we showed that the $q^{2}\beta_{e}=C_{\mathrm{nl}}$ criterion~\cite{zhang2026a,zhang2026b} predicts the critical value of $q^{2}\beta_{e}$ for the onset of large electromagnetic turbulence in simulations of STEP-EC-HD. We were also able to extend it to explore saturation at larger $q^{2}\beta_{e}.$ An important question is to what extent this theory can be useful in the design of future STEP operating points. Calculating the critical value of the parameter \( q^{2} \beta_{e} \) still requires a nonlinear gyrokinetic calculation per local equilibrium. Although these simulations can be performed at lower values of $q^{2}\beta_{e},$ they are still relatively expensive in comparison to, e.g., quasilinear modelling. Using this theory as part of integrated modelling is still prohibitively expensive over confinement and resistive timescales, particularly in modelling ramp-up and ramp-down where the global equilibrium is also evolving in time. 

\subsection{Connecting the stress-balance picture to ideal ballooning stability}

In the simulations presented in Section~\ref{sec:numerical_simulations}, $\beta^\prime$ was scaled proportionally to $\beta_{e}$. As a result, the dimensionless parameter $q^2 \beta_e$ is closely related to the ideal MHD ballooning stability parameter $\alpha_{\mathrm{MHD}} = -R_0 q^2 \mathrm{d}\beta / \mathrm{d}r$, which strongly influences the competition between pressure-gradient drive and magnetic field-line bending. The onset of the ideal ballooning mode (IBM) (see e.g., \cite{davies2022}) corresponds to the critical point at which this balance is lost and pressure-driven field-aligned Alfv\'enic instabilities emerge. This same parameter appears in gyrokinetic formulations of pressure-gradient-driven modes such as the kinetic ballooning mode and hKBM, indicating a connection between ideal MHD stability and gyrokinetic turbulence dynamics. As such, the IBM boundary in $(q, \beta_e)$ or $(\hat{s}, \beta_e)$ space could offer a natural proxy for identifying regions of parameter space where the nonlinear saturation mechanisms of conventional electrostatic turbulence  may fail. This idea was briefly discussed in \cite{kennedy2023a}; where it was shown that the hKBM instability in STEP-EC-HD tracks the behaviour of the IBM. 

Although the IBM is a linear MHD instability, its physics is very relevant to nonlinear gyrokinetic calculations of Alfv\'enic turbulence. In the simulations presented in Section~\ref{sec:numerical_simulations}, the transition to a high-$\beta_e$, low-zonal-shear state does not correspond to a sharp change in linear drive; rather, it occurs due to a reorganisation of the nonlinear stress dynamics. When stabilisation from field-line bending is limited, at higher $\beta_e$ magnetic flutter and its associated Maxwell stresses become stronger and this starts to couple efficiently into the zonal-flow momentum balance, undermining electrostatic saturation and enabling the transition to a non-zonal, high-transport state. Increasing $\beta_e$ generates a boundary in parameter space beyond which the nonlinear saturation pathway changes, not because of stronger linear growth, but because the character of the nonlinear stress balance is fundamentally altered. This suggests that IBM stability can act as a valuable proxy for anticipating the breakdown of zonal-flow-regulated turbulence. As the turbulence becomes increasingly electromagnetic, it triggers the onset of a regime in which the standard saturation mechanism of electrostatic turbulence from zonal flows is greatly weakened, transport fluxes are boosted, and different saturation mechanisms, or a transition to global or MHD-like behaviour, may be required.

\subsection{Ideal ballooning eigenvalue as a stability metric}

IBM instability arises when the destabilising pressure gradient in regions of ``bad'' magnetic curvature, defined by \( (\mathbf{b} \cdot \nabla \mathbf{b}) \cdot \nabla p > 0 \), overcomes the stabilising influence of field-line bending. The relevant perturbations are ballooning in nature: they are localised along magnetic field lines but tend to bulge outwards in regions of unfavourable curvature, where the pressure gradient is steepest. The stability of the plasma to such perturbations is governed by the ballooning mode equation \cite{connor1979,dewar1983}, which  is a second-order one-dimensional eigenvalue problem for the perturbation \( X_b(\vartheta) \):
\begin{equation}
\frac{1}{J} \frac{d}{d\vartheta} \left( \frac{|\nabla \alpha|^2}{J B^2} \frac{dX_b}{d\vartheta} \right)
+ 2 \frac{dp}{d\psi}
\left( \frac{\mathbf{B} \times \nabla(\mu_0 p + B^2/2)}{B^2} \cdot \nabla \alpha \right) X_b
= -\rho \omega^2 \frac{|\nabla \alpha|^2}{B^2} X_b,
\label{eq:ibm}
\end{equation}
subject to the boundary condition
\begin{equation}
\lim_{\vartheta \to \pm \infty} X_b(\vartheta; \psi, \alpha, \vartheta_0) = 0.
\label{eq:ibm_bc}
\end{equation}
Here, \( \vartheta \) is the ballooning angle, \( \vartheta_0 \) the ballooning parameter, \( \psi \) the flux-surface label, and \( \rho \) the plasma mass density. The Jacobian is defined as \( J = (\mathbf{B} \cdot \nabla \vartheta)^{-1} \), and \( \alpha \) is the field-line label. The boundary condition \( \lim_{\vartheta \to \pm \infty} X_b(\vartheta) = 0 \) ensures a well-posed eigenvalue problem, typically leading to a discrete, ordered spectrum of \( \omega^2 \), with the lowest eigenvalue corresponding to the most unstable ballooning perturbation. The eigenvalue \(\hat{\lambda} \equiv -\omega^2 \) determines the mode stability. 


\subsection{Ideal ballooning calculations}

A key advantage of the IBM formalism is that it allows efficient scans over equilibrium parameters \cite{Gaur_Buller_Ruth_Landreman_Abel_Dorland_2023}. In Figure~\ref{fig:IBM-contours}, we use the ballooning-mode diagnostic in \texttt{pyrokinetics} to calculate the IBM eigenvalue, $\hat{\lambda} \equiv -\omega^{2}$, in the $(q,\beta_{e})$ (top) and $(\hat{s},\alpha_{\mathrm{MHD}})$ (bottom) planes for STEP-EC-HD (left) and CBC (right). Positive values indicate unstable growth rates, while negative values correspond to stable oscillations. The dashed black lines mark the IBM stability boundary, defined by the change in sign of the most unstable eigenvalue. The solid black lines denote the nonlinear transition condition, $q^{2}\beta_{e} = C_{\mathrm{nl}}$. In both STEP-EC-HD and CBC, the nonlinear transition lies well below the IBM threshold, confirming that the transition to high fluxes occurs before the onset of ideal MHD instability (see Section~\ref{sec:numerical_simulations}). In the $(q,\beta_{e})$ stability map of Figure~\ref{fig:IBM-contours} the IBM boundary lies well above the $q^{2}\beta_{e}=C_{\mathrm{nl}}$ transition, approaching it only at small $q$, so the link between ideal ballooning and the transport is not that the two boundaries lie on top of one another. Rather, the \emph{isolines} of $\hat{\lambda}$ track the regions of large heat flux across the whole plane (Figure~\ref{fig:q-beta-step-IBM}), tying the ballooning stability limit to the underlying Alfv\'enic physics. The IBM boundary is thus best read as a cheap \emph{proxy} that lies above, and tracks, the nonlinear transition. The black star marks the location of the STEP-EC-HD flat-top point, while the red star marks a second-stable operating point in CBC (discussed in Section~\ref{subsec:reverse_transition}).

\begin{figure}
  \centering
  \begin{minipage}[t]{0.5\textwidth}
    \centering
    \includegraphics[]{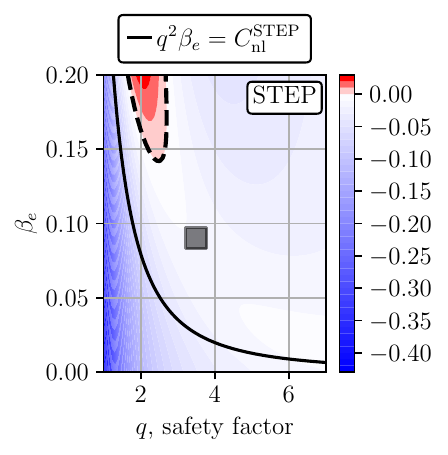}
  \end{minipage}\hfill
  \begin{minipage}[t]{0.5\textwidth}
    \centering
    \includegraphics[]{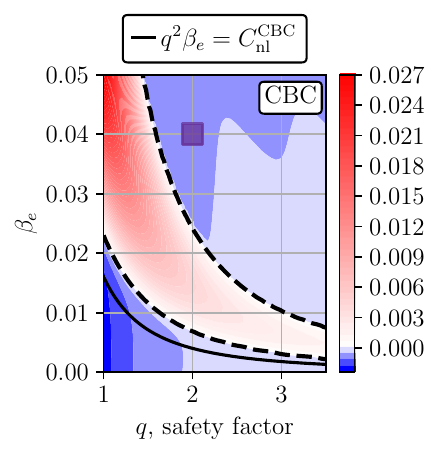}
  \end{minipage}
\\
  \centering
  \begin{minipage}[t]{0.5\textwidth}
    \centering
    \includegraphics[]{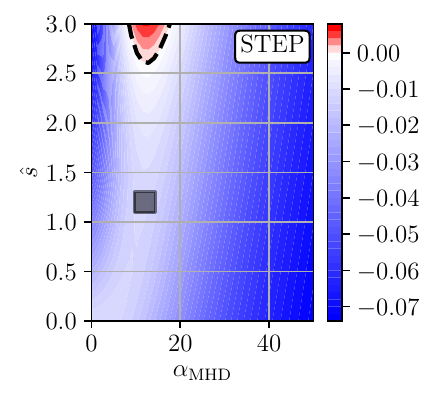}
  \end{minipage}\hfill
  \begin{minipage}[t]{0.5\textwidth}
    \centering
    \includegraphics[]{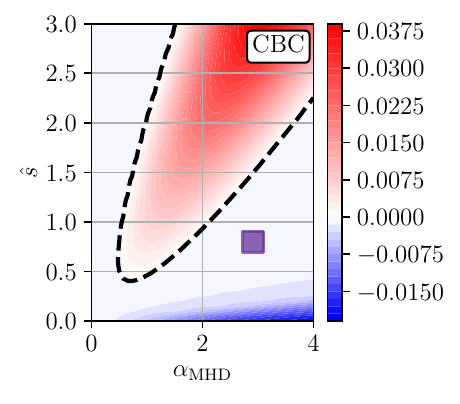}
  \end{minipage}

  \caption{Ideal ballooning mode (IBM) eigenvalue, $\hat{\lambda} \equiv -\omega^{2}$ in the $(q,\beta_{e})$ plane (top row) and  $(\hat{s},\alpha_{\mathrm{MHD}})$ plane (bottom row) of STEP-EC-HD (left-hand side) and CBC (right-hand side). Positive values denote growth rates (instability) and negative values denote real frequencies (stable waves). The dashed black line shows the IBM stability boundary. The solid black line shows the nonlinear transition boundary $q^{2}\beta_{e} = C_{\mathrm{nl}}.$ A black square marks the position of the STEP-EC-HD flat-top. A purple square marks the location of a second-stable point for CBC (see Section~\ref{subsec:reverse_transition}).} 
  \label{fig:IBM-contours}
\end{figure}

The purple box in Figure~\ref{fig:IBM-contours} corresponds to the high-$\beta_{e}$ nonlinear CBC simulation examined in Section~\ref{subsec:reverse_transition}. This point falls within the so-called second-stable region, beyond the IBM-unstable domain, illustrating the connection between reduced flux levels and IBM physics at high $\beta_{e}$. In the $(q,\beta_{e})$ plane, the extent of the IBM-unstable region is determined by the flux-surface geometry, suggesting that shaping could provide a continuous path between the first- and second-stable regions without entering the region associated with large gyrokinetic fluxes. 

This connection is further illustrated in Figure~\ref{fig:q-beta-step-IBM}, which overlays the IBM eigenvalue $\hat{\lambda}$ on top of the nonlinear simulation results reported in Figure~\ref{fig:q-beta-step}: blue circles indicate simulations converging to low fluxes ($Q_{\mathrm{TOT}}/Q_{\mathrm{NZT}} < 1$), red crosses indicate simulations that remain in the high-transport state, and contours of $\hat{\lambda}$ emphasise the connection between the low-flow, high-$q^{2}\beta_{e}$ state and the second-stable region (i.e., blue circles are located at more negative values of $\hat{\lambda}$). While $\hat{\lambda}$ also correlates with the first transition to large fluxes, its predictive power is weaker in that regime since the transition is intrinsically nonlinear.

\begin{figure}
  \centering
\includegraphics[width=0.75\textwidth]{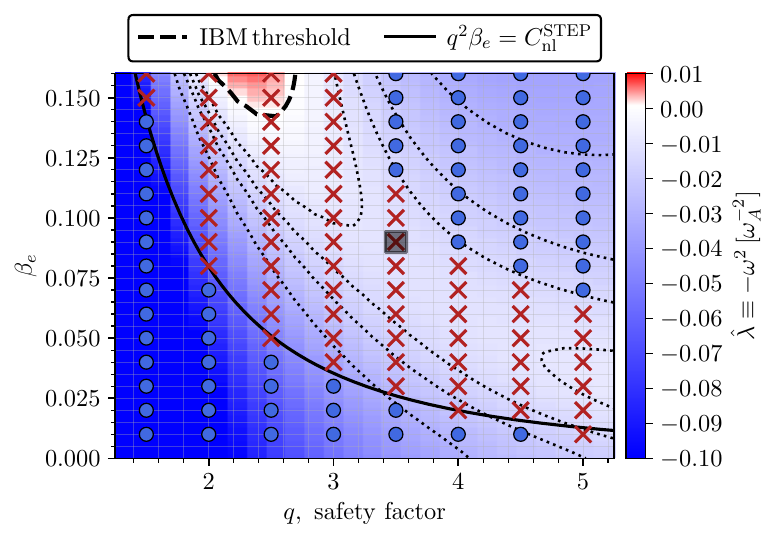}
  \caption{Results from nonlinear simulations of STEP-EC-HD at various values of $q$ and $\beta_{e}$ using \texttt{stella}. Other than $\beta^{\prime},$ which is always proportional to $\beta_{e},$ all other local parameters are fixed. Simulations which converge to $Q_{\mathrm{TOT}}/Q_{\mathrm{NZT}} < 1$ are marked by \textcolor{blue}{blue circles}. Simulations which do not converge to $Q_{\mathrm{TOT}}/Q_{\mathrm{NZT}} < 1$ are marked by \textcolor{red}{red crosses}. The dashed black line (top left) is the ideal ballooning mode threshold calculated using \texttt{pyrokinetics}. The solid black line is the predicted transition curve $q^{2}\beta_{e} = C_{\mathrm{nl}}^{\mathrm{STEP}}$ \cite{zhang2026b} where $C_{\mathrm{nl}}^{\mathrm{STEP}}$ is computed by fitting a single simulation at $(q, \beta_{e}) = (3.5,0.026)$. The black box is the nominal STEP-EC-HD operating point. The colour scale shows the IBM eigenvalue $\hat{\lambda} \equiv -\omega^{2}$. Positive values denote growth rates (instability) and negative values denote real frequencies (stable waves). Dotted lines are isolines of the IBM eigenvalue.}
  \label{fig:q-beta-step-IBM}
\end{figure}

Connections between IBM and stress-balance are summarised in Table~\ref{tab:stressbalance_ibm}.

\begin{table}
    \centering
    \renewcommand{\arraystretch}{1.2}
    \begin{tabular}{p{5cm} p{4.5cm} p{4.5cm}}
        \toprule
        \textbf{Observation} & \textbf{Stress balance of \cite{zhang2026b}} & \textbf{Ideal Ballooning} \\
        \midrule
        Forward transition (Section~\ref{subsec:forward_transition}) &
        $q^{2}\beta_{e} = C_{\mathrm{nl}}^{\mathrm{STEP}}$ \newline Net-negative zonal energy transfer &
        Proximity to IBM boundary \newline Onset of ideal-pressure-driven Alfv\'enic modes \\
        
        Reverse transition (Section~\ref{subsec:reverse_transition}) &
        Recovery of zonal shear via restructured nonlinear stress \newline Suppression of hKBM &
        Access to second-stable region \newline Ballooning modes suppressed; IBM stable \\
        
        hKBM instability (cf. \cite{kennedy2023a}) &
        $\Pi_{\mathrm{lin}}$ dominated by curvature-coupled terms \newline Enhanced sensitivity to geometry and $\hat{s}$ &
        Tracks IBM in $k_y \rho_s \ll 1$ limit~\cite{Tang1980} \\
        
        Proximity to IBM boundary &
        Indicates structural shift in nonlinear stress components \newline Stress balance vulnerable to breakdown &
        Identifies transition between stable and unstable ideal pressure-gradient response \\
        \bottomrule
    \end{tabular}
    \caption{Correspondence between nonlinear gyrokinetic observations and ideal ballooning stability. The IBM boundary acts as a proxy for where zonal-flow-based saturation fails, either due to overwhelming linear drive or to changes in the nonlinear stress structure.}
    \label{tab:stressbalance_ibm}
\end{table}

\section{Conclusions}
\label{sec:conclusions} 

In this work, we have investigated the nonlinear transition to large fluxes found in local gyrokinetic simulations of electromagnetic turbulence in STEP-like spherical tokamak plasmas \cite{Giacomin2024}. Across more than 100 nonlinear gyrokinetic simulations we have identified the critical value of $q^2\beta_e$ at which STEP-EC-HD turbulence runs away, and shown that the stress-balance framework of~\cite{zhang2026a,zhang2026b} accounts for it in STEP. This is the critical value of $q^2 \beta_e$ beyond which (without flow shear) local gyrokinetic predictions of the transport flux in STEP-EC-HD undergo a transition from a low-flux regime to an EM high-transport regime; moreover, this transition arises with no obvious qualitative change in the linear stability spectrum, and in conditions stable to ideal $n=\infty$ ballooning modes. We showed that this transition is governed by a delicate interplay between $\bm{E} \times \bm{B}$ and magnetic-flutter stresses. Linear physics can enable the system to enter a second-stable regime at sufficiently high $\beta^{\prime}$ (and $\beta_e$). Our results reveal that local simulations can exhibit both a stable saturated state and an electromagnetic large-transport state depending on the choice of $\beta_e$, suggesting the presence of a bifurcation structure in turbulent transport. This is reminiscent of “second-stability” phenomenology long discussed in MHD ballooning theory and in modern pedestal and internal-transport-barrier modelling, where second stability physics can give access to moderate transport fluxes at high gradients. Finally, because the threshold scales approximately as $\beta_{e,\mathrm{crit}}\sim L_T/(q^2R)$, the same MHD-controlled saturation constraint is accessed at lower $\beta_e$ in conventional-aspect-ratio tokamaks (assuming $q$, $a$ and $L_{T}$ remain constant), motivating future work to assess whether the STEP-like no-go regions identified here connect systematically to transport-barrier physics across aspect ratio.

These findings have important implications for future flux-driven simulations. Unlike conventional turbulence studies with fixed background gradients, flux-driven simulations allow the pressure gradient to evolve self-consistently, offering a more faithful description of the underlying plasma dynamics. However, the presence of a transport bifurcation introduces sensitive dependence on initial conditions and modelling assumptions. In flux-driven simulations of STEP-EC-HD it is possible to observe two qualitatively distinct outcomes: either the mean pressure gradient relaxes to a near-marginal state, resulting in weak confinement and inadequate fusion power, or the system crosses into a second-stable regime, where increasing pressure gradients generate a more favourable confinement regime with strong fusion performance (both of these behaviours have been observed in flux-driven simulations \cite{Giacomin_Dickinson_Dorland_Mandell_Bokshi_Casson_Dudding_Kennedy_Patel_Roach_2025}). Schematically, which outcome is realised depends on whether the self-consistently evolving $\beta'$ is driven through the high-transport band into the second-stable region (favourable) or instead stalls below it near marginal stability (unfavourable), and hence on the available heating power, the current ($q$) profile, and any residual flow shear. This highlights a crucial role for theory in pre-screening operational scenarios and guiding expensive flux-driven simulations towards physically more favourable regions of parameter space. Predictive integrated modelling for scenario design will require identifying the conditions that mitigate large electromagnetic fluxes at high~$\beta_e$. This paper provides a criterion for the onset of such fluxes, which can then be used to assess the potential for mitigation, e.g., through strong~$\beta'$, negative magnetic shear, or sufficient flow shear.

The role of flow shear in the nonlinear saturation of electromagnetic turbulence is another key consideration for STEP. In present-day experiments, such as MAST-U and ST40, strong equilibrium flow shear arises naturally due to neutral beam injection and plays a major role in suppressing turbulent transport. However, STEP is expected to operate with low external torque and correspondingly weak background shear. It stands to reason that the streamer-like structures responsible for electromagnetic high-transport states are relatively insensitive to whether they are sheared by zonal flows or by radial gradients in the equilibrium rotation profile; what matters is the presence of sufficient shear to decorrelate them. Large transport regimes can be expected in plasma conditions where equilibrium flow shear is weak, and turbulent stresses resist the generation of zonal flows, though this will be mitigated in plasma regions with negative magnetic shear and/or high $\beta^{\prime}$. Going forward, a key challenge for fusion power plant (FPP) concepts like STEP will be to exploit high fidelity transport models to explore parameter space and to seek robust viable routes to access burning plasma regimes, using the available actuators. More detailed studies are required to determine whether tailoring the current profile, or generating modest levels of flow shear—either through heating and current drive, intrinsic rotation, or other experimental means—can assist in this endeavour.

Our stress-balance analysis also connects to the broader theory of zonal-flow generation. The Reynolds ($\bm{E}\times\bm{B}$) versus Maxwell (magnetic-flutter) competition examined here is a distinct axis from the established partition of the electrostatic zonal drive into beat-driven and spontaneous (modulational) contributions~\cite{Chen2024}: the Maxwell stress opposes the $\bm{E}\times\bm{B}$ drive irrespective of whether the latter is beat-driven or modulationally excited, so the non-zonal transition is most naturally understood as the magnetic-flutter stress overcoming the zonal-flow drive (an approach to the ideal-Alfv\'enic near-cancellation) rather than as a loss of modulational instability. In principle the time-resolved zonal torques (Figures~\ref{fig:STEP_beta0p020_fig2},~\ref{fig:STEP_beta0p030_fig3} and~\ref{fig:CBC_highbeta_secondstability_fig1}) could discriminate these mechanisms, a beat-driven drive tracking the primary-fluctuation envelope whilst a spontaneous drive exhibits a delayed, self-amplifying transient, although a systematic separation is beyond the scope of the present work. A related open question is the role of phase-space zonal structures (PSZS)~\cite{ChenN2026}, the velocity-space-resolved $n=0$ response that is not captured by the zonal field potentials alone. PSZS are a generic feature of nonlinear gyrokinetics and are therefore present in our simulations, but our analysis targets the field-moment (stress) balance rather than the phase-space response, and we have not attempted to isolate them. Whether PSZS contribute materially to hKBM saturation in STEP remains to be established; using velocity-space-resolved diagnostics, such as those found in \texttt{stella}, one could in principle support such an analysis by comparing the flux-surface-averaged gyrocentre response against the zonal fields, and we leave this to future work.

In summary, our results have combined nonlinear local gyrokinetic simulations with theoretical insights to improve our understanding of how electromagnetic turbulence is likely to affect transport in high $\beta_e$ FPP concepts.  This work will contribute to the development of higher fidelity transport models that will be essential to constrain the operational space and guide access to high performance in future fusion devices. 

\section*{Acknowledgements}

The authors would like to thank A.A. Schekochihin, M. Barnes, M. Hardman, R. Nies, G. Merlo, D. Dickinson, A. Bokshi, and M. Giacomin. This work has been
funded by STEP, a major technology and infrastructure programme led by UK Industrial Fusion Solutions Ltd (UKIFS),
which aims to deliver the UK’s prototype fusion power plant
and a path to the commercial viability of fusion. Part of this work was performed using resources provided by the PITAGORA supercomputer from the National Supercomputing
Consortium CINECA, under the projects GEMST24 and GEMST25.  Part of this work was performed using resources provided by the Cambridge Service for Data Driven Discovery (CSD3) operated by the University of Cambridge Research Computing Service (\url{www.csd3.cam.ac.uk}), provided by Dell EMC and Intel using Tier-2 funding from the Engineering and Physical Sciences Research Council (capital grant EP/T022159/1), and DiRAC funding from the Science and Technology Facilities Council (\url{www.dirac.ac.uk}). The work of T.A. was supported in part by the Laboratory Directed Research and Development
(LDRD) Program at the Princeton Plasma Physics Laboratory for the U.S. Department of Energy under Contract
No. DE-AC02-09CH11466. The United States Government retains a non-exclusive, paid-up, irrevocable, world-wide
license to publish or reproduce the published form of this manuscript, or allow others to do so, for United States
Government purposes.

\clearpage

\section*{References}
\bibliographystyle{unsrt}
\bibliography{bibliography}

 \appendix

\section{The $\delta\!B_\parallel$ contribution to the zonal-flow energy}
\label{app:zonal_energy_budget}

Equation~(\ref{eq:zonal_evolution_equation}) for the zonal potential (vorticity) retains the parallel magnetic-compression ($\delta\!B_\parallel$) part of the generalised polarisation, whereas the main-text energy budget~(\ref{eq:zonal_energy_balance}) neglects it. In this appendix we (i) order this term analytically, (ii) derive the zonal-flow energy balance~(\ref{eq:zonal_energy_balance}) from~(\ref{eq:zonal_evolution_equation}), keeping the $\delta\!B_\parallel$ exchange term explicit, and (iii) quantify, directly from the nonlinear simulations of Section~\ref{sec:numerical_simulations}, the $\delta\!B_\parallel$ contributions to the zonal-flow \emph{energy}. We find that, while the compressive turbulent \emph{stress} $\Pi_{\delta\!B_\parallel}$ (which is retained throughout this work) is not small, the $\delta\!B_\parallel$ contribution to the zonal-flow \emph{energy} and \emph{polarisation} is small across the entire set of simulations in this paper, which bracket both the forward (non-zonal) and reverse transitions.

\subsection{Ordering of the parallel magnetic compression}
\label{app:bpar_ordering}

The size of $\delta\!B_\parallel$ relative to the electrostatic response follows from the perpendicular part of Amp\`ere's law~(\ref{eq:ampere_perpendicular}), which expresses the perpendicular pressure balance of the fluctuations. Ordering the perturbed distribution to support a Boltzmann-like response, with comparable perpendicular-pressure perturbations,
\begin{equation}
\frac{h_s}{F_s}\sim\frac{\delta n_s}{n_s}\sim\frac{q_s\phi}{T_s}\sim\frac{\delta p_{\perp s}}{p_s},
\label{eq:app_bpar_orderings}
\end{equation}
perpendicular pressure balance gives
\begin{equation}
\frac{\delta\!B_\parallel}{B}\;\sim\;\beta_s\,\frac{\delta p_{\perp s}}{p_s}\;\sim\;\beta_s\,\frac{q_s\phi}{T_s}
\qquad\Longrightarrow\qquad
\frac{\delta\!B_\parallel}{B}\;\ll\;\frac{q_s\phi}{T_s},
\label{eq:app_bpar_ordering}
\end{equation}
where $\beta_s = 8\pi n_s T_s/B^2$ and the inequality holds because $\beta_s\ll1$ for the cases of interest [$\beta_e\lesssim O(10\%)$]. The parallel magnetic compression entering the generalised polarisation is therefore suppressed by one power of $\beta$ relative to the electrostatic polarisation, and the associated compressive \emph{energy}, being quadratic in $\delta\!B_\parallel$, is suppressed by $O(\beta^2)$. This is an ordering, not a demonstration that the energy \emph{transfer} is negligible; the latter is established quantitatively in~\ref{app:bpar_numerics}.

\subsection{Derivation of the zonal-flow energy balance}
\label{app:energy_derivation}

Here we derive the zonal-flow energy balance~(\ref{eq:zonal_energy_balance}) explicitly from~(\ref{eq:zonal_evolution_equation}), the flux-surface-averaged evolution equation for the generalised zonal polarisation charge, retaining the $\delta\!B_\parallel$ polarisation throughout. The flux-surface average in~(\ref{eq:zonal_evolution_equation}) annihilates all non-zonal ($k_y\neq0$) components and retains only the zonal ($k_y=0$) modes, for which $\bm{k}_\perp=k_x\nabla x.$ Taking a single such mode and writing out the species sums in the generalised polarisation charge, the vorticity equation reads, per mode,
\begin{equation}
\frac{\partial}{\partial t}\left[\sum_s \frac{q_s^2 n_s}{T_s}\bigl(1-\Gamma_{0s}\bigr)\langle\phi\rangle_{\psi,k_x}
- \sum_s q_s n_s\,\Gamma_{1s}\left\langle\frac{\delta\!B_\parallel}{B}\right\rangle_{\psi,k_x}\right]
= \Pi_{\mathrm{lin},k_x}+\Pi_{\phi,k_x}+\Pi_{A_\parallel,k_x}+\Pi_{\delta\!B_\parallel,k_x},
\label{eq:app_scalar}
\end{equation}
where all symbols are defined as in the main text.

To form an energy, we project~(\ref{eq:app_scalar}) onto the zonal potential. We multiply through by the complex conjugate $\langle\phi\rangle^*_{\psi,k_x}$ and take the real part. The finite-Larmor-radius weights are real and independent of time, so they commute with both taking the real part and the time derivative. For the electrostatic polarisation term we use the identity $\mathrm{Re}(z^*\,\partial_t z)=\tfrac12\,\partial_t|z|^2$ with $z=\langle\phi\rangle_{\psi,k_x}$, so that
\begin{equation}
\mathrm{Re}\!\left[\langle\phi\rangle^*_{\psi,k_x}\,\frac{\partial}{\partial t}\sum_s \frac{q_s^2 n_s}{T_s}\bigl(1-\Gamma_{0s}\bigr)\langle\phi\rangle_{\psi,k_x}\right]
= \frac{\partial}{\partial t}\left[\frac12\sum_s \frac{q_s^2 n_s}{T_s}\bigl(1-\Gamma_{0s}\bigr)\bigl|\langle\phi\rangle_{\psi,k_x}\bigr|^2\right].
\label{eq:app_phi_quadratic}
\end{equation}

The parallel-compression polarisation term, by contrast, couples the two distinct fields $\langle\phi\rangle_{\psi,k_x}$ and $\langle\delta\!B_\parallel/B\rangle_{\psi,k_x}$, and so \emph{cannot} be written as the time derivative of a quadratic in a single field; it survives the projection as a genuine cross term that we define as the exchange torque,
\begin{equation}
T_{\mathrm{exch},k_x} \equiv \sum_s q_s n_s\,\Gamma_{1s}\,
\mathrm{Re}\!\left(\langle\phi\rangle^*_{\psi,k_x}\,\frac{\partial}{\partial t}\left\langle\frac{\delta\!B_\parallel}{B}\right\rangle_{\psi,k_x}\right).
\label{eq:exchange_transfer}
\end{equation}

Projecting the right-hand side of~(\ref{eq:app_scalar}) in the same way reproduces the four zonal torques of~(\ref{eq:lin_transfer})--(\ref{eq:bpar_transfer}),
\begin{equation}
\mathrm{Re}\!\left[\langle\phi\rangle^*_{\psi,k_x}\bigl(\Pi_{\mathrm{lin},k_x}+\Pi_{\phi,k_x}+\Pi_{A_\parallel,k_x}+\Pi_{\delta\!B_\parallel,k_x}\bigr)\right]
= T_{\mathrm{lin},k_x}+T_{\phi,k_x}+T_{A_\parallel,k_x}+T_{\delta\!B_\parallel,k_x}.
\label{eq:app_rhs_proj}
\end{equation}

Collecting~(\ref{eq:app_phi_quadratic})--(\ref{eq:app_rhs_proj}) (and moving the cross term to the right-hand side) gives, with no further approximation, the per-mode zonal-flow energy balance,
\begin{equation}
\frac{\partial}{\partial t}\left[\frac12\sum_s \frac{q_s^2 n_s}{T_s}\bigl(1-\Gamma_{0s}\bigr)\bigl|\langle\phi\rangle_{\psi,k_x}\bigr|^2\right]
= T_{\mathrm{lin},k_x}+T_{\phi,k_x}+T_{A_\parallel,k_x}+T_{\delta\!B_\parallel,k_x}+T_{\mathrm{exch},k_x}.
\label{eq:app_energy_full}
\end{equation}

It remains to express the polarisation energy on the left-hand side of~(\ref{eq:app_energy_full}) in closed form. We specialise to a plasma of main ions of charge $q_i=Ze$ and electrons; the electron polarisation is smaller than the ion polarisation by $O(m_e/m_i)$ and is dropped. In the long-wavelength limit $\alpha_i\to0$ we have $1-\Gamma_{0i}\simeq\alpha_i=\tfrac12 k_x^2|\nabla x|^2\rho_i^2$, so the ion polarisation weight reduces to
\begin{equation}
\frac12\,\frac{q_i^2 n_i}{T_i}\,\bigl(1-\Gamma_{0i}\bigr)
\;\simeq\; \frac{(Ze)^2 n_i}{4\,T_i}\,k_x^2|\nabla x|^2\rho_i^2
\;=\; \frac{n_i m_i c^2}{2 B^2}\,k_x^2|\nabla x|^2,
\label{eq:app_EZF_coeff}
\end{equation}
where the last equality substitutes $\rho_i^2=v_{\mathrm{th}i}^2/\Omega_i^2=2 T_i m_i c^2/(Z^2 e^2 B^2)$, so that the factors of $Z$, $e$ and $T_i$ cancel. The zonal-flow polarisation energy is therefore
\begin{equation}
E_{\mathrm{ZF}}(k_x)\;=\;\frac{n_i m_i c^2}{2 B^2}\,k_x^2|\nabla x|^2\,\bigl|\langle\phi\rangle_{\psi,k_x}\bigr|^2,
\label{eq:app_EZF}
\end{equation}
with effective inertia set by the ion polarisation density $n_i m_i c^2/B^2$. With this identification, the long-wavelength limit of the per-mode balance~(\ref{eq:app_energy_full}) has the form of the zonal-flow energy balance~(\ref{eq:zonal_energy_balance}) used in the main text, with the single exchange term $T_{\mathrm{exch}}$ added on the right-hand side. The main-text balance~(\ref{eq:zonal_energy_balance}) quotes this same prefactor; substituting $\rho_i^2$ then gives the fully reduced form~(\ref{eq:app_EZF}).

Beyond the long-wavelength reduction of the left-hand side, the only term separating the per-mode balance~(\ref{eq:app_energy_full}) from the main-text balance~(\ref{eq:zonal_energy_balance}) is the exchange torque $T_{\mathrm{exch}}$, which we now drop. This, and the size of the compressive zonal energy
\begin{equation}
E_{\delta\!B_\parallel}(k_x) \equiv \frac12\sum_s q_s n_s\,\Gamma_{1s}\left|\left\langle\frac{\delta\!B_\parallel}{B}\right\rangle_{\psi,k_x}\right|^2
\label{eq:app_Edbpar}
\end{equation}
relative to $E_{\mathrm{ZF}}$, are assessed numerically in~\ref{app:bpar_numerics}. We note that $\delta\!B_\parallel$ is not an independent dynamical field: it is fixed instantaneously by the perpendicular Amp\`ere's law~(\ref{eq:ampere_perpendicular}), i.e.,\ slaved to $h_s$ through perpendicular pressure balance, and so is not a separate dynamical energy reservoir. The zonal-flow energy considered here is the $\phi$-based $E_{\mathrm{ZF}}$ of~(\ref{eq:app_EZF}); the parallel magnetic compression enters it only through (a) the exchange torque $T_{\mathrm{exch}}$ of~(\ref{eq:exchange_transfer}) and (b) the compressive turbulent stress $\Pi_{\delta\!B_\parallel}$ of~(\ref{eq:stress_compressive}), which is retained on the right-hand side. 

Note that a combined two-field quadratic energy,
\begin{equation}
\frac12\sum_s \frac{q_s^2 n_s}{T_s}\bigl(1-\Gamma_{0s}\bigr)\bigl|\langle\phi\rangle_{\psi,k_x}\bigr|^2
+ \frac12\sum_s q_s n_s\,\Gamma_{1s}\left|\left\langle\frac{\delta\!B_\parallel}{B}\right\rangle_{\psi,k_x}\right|^2,
\label{eq:app_twofield}
\end{equation}
is not uniquely defined (the cross drive may be partitioned between the two fields in more than one way) and is unnecessary once $T_{\mathrm{exch}}$ and $E_{\delta\!B_\parallel}$ are found to be small, as we now show.

\subsection{Numerical assessment}
\label{app:bpar_numerics}

We evaluate, directly from the nonlinear \texttt{stella} simulations and as functions of $k_x$ and time, the compressive zonal energy $E_{\delta\!B_\parallel}$ of~(\ref{eq:app_Edbpar}) relative to the zonal-flow energy $E_{\mathrm{ZF}}$ of~(\ref{eq:app_EZF}), and the exchange torque $T_{\mathrm{exch}}$ of~(\ref{eq:exchange_transfer}) relative to the dominant turbulent torques. Sums are taken over the large-scale zonal modes (below the heat-flux-spectrum peak) and averages over the last 10\% of each simulation.

Figure~\ref{fig:bpar_assessment} summarises the same set of simulations shown in Figure~\ref{fig:q-beta-step-transition-partitioned}, organised into the same three physically distinct regimes: the pre-transition regime [panels (a), (d), (g), and (j)]; the electromagnetic transition [panels (b), (e), (h), and (k)]; and the post-transition regime [panels (c), (f), (i), and (l)]. The compressive zonal energy [panels (a)--(c)] follows the ordering of~(\ref{eq:app_bpar_ordering}), $E_{\delta\!B_\parallel}/E_{\mathrm{ZF}}=O(\beta^2)$ and never exceeds $\approx6\times10^{-2}$, while the exchange torque $\mathcal{S}(T_{\mathrm{exch}})$ [panels (d)--(f)] stays at the few-percent level at all $q^2\beta_e$ and in every regime, where the time-integrated ``size'' of a quantity $X$ is
\begin{equation}
\mathcal{S}(X) \;\equiv\; \frac{\int \mathrm{d} t \, |X|}{\int \mathrm{d} t \, \!\left(|T_\phi|+|T_{A_\parallel}|\right)}.
\label{eq:size_measure}
\end{equation} In contrast, the compressive turbulent \emph{stress} $\Pi_{\delta\!B_\parallel}$ is retained throughout and is not small: $\mathcal{S}(T_{\delta\!B_\parallel})$ [panels (g)--(i)] is an $O(0.1$--$1)$ fraction, so the parallel magnetic compression is fully accounted for where it matters, namely as a turbulent stress on the right-hand side of~(\ref{eq:zonal_energy_balance}). 

\begin{figure}
\centering
\includegraphics[width=\textwidth]{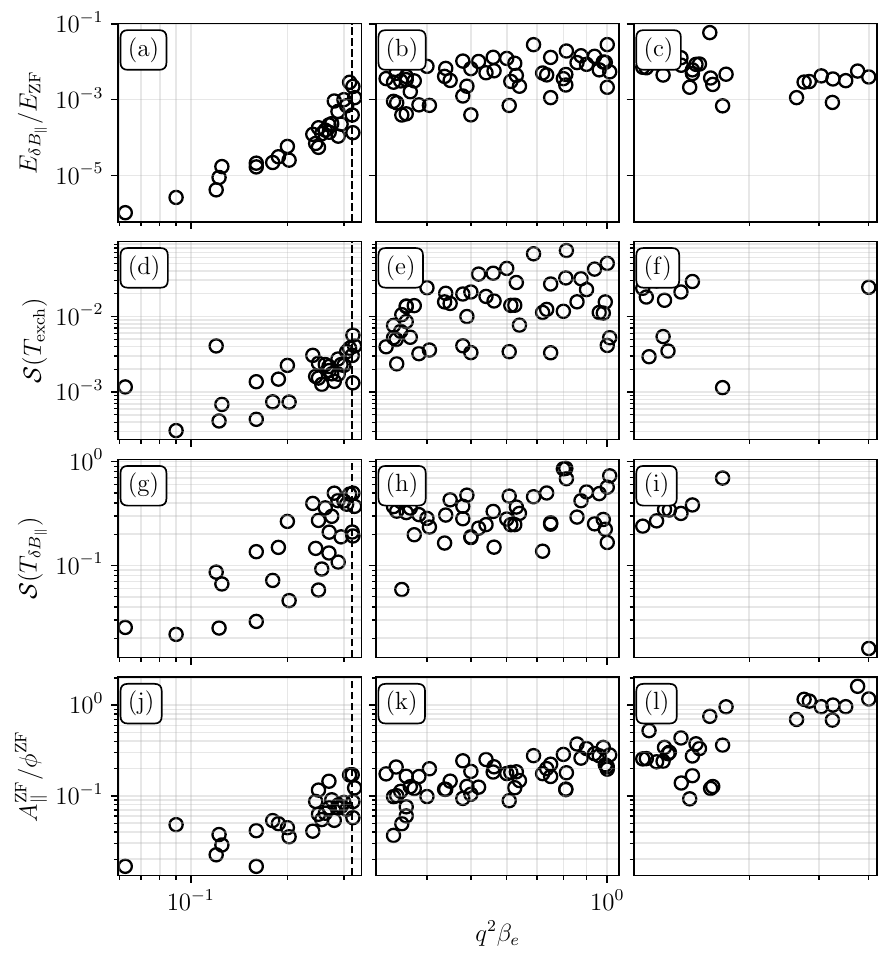}
\caption{Parallel-magnetic-compression diagnostics as a function of $q^{2}\beta_{e}$ for the data points in Figure~\ref{fig:q-beta-step-transition-partitioned}: the compressive zonal energy $E_{\delta\!B_\parallel}/E_{\mathrm{ZF}}$ (top row); the exchange-torque size $\mathcal{S}(T_{\mathrm{exch}})$ [Equation~(\ref{eq:size_measure})] (second row); the compressive-stress size $\mathcal{S}(T_{\delta\!B_\parallel})$ (third row); and the zonal parallel-current amplitude $A_\parallel^{\mathrm{ZF}}/\phi^{\mathrm{ZF}}$ (bottom row). The black dashed vertical line is $q^{2}\beta_{e} = C_{\mathrm{nl}}^{\mathrm{STEP}}.$ Each column corresponds to a physically distinct regime: before the non-zonal transition [(a), (d), (g), and (j)]; during the electromagnetic transition [(b), (e), (h), and (k)]; and after the reverse transition [(c), (f), (i), and (l)].}
\label{fig:bpar_assessment}
\end{figure}

In the cases examined, the parallel magnetic compression therefore enters the zonal-flow energy only as a small exchange, consistent with $\delta\!B_\parallel$ being fixed instantaneously by perpendicular Amp\`ere's law~(\ref{eq:ampere_perpendicular}) rather than evolving as an independent dynamical field. This supports working with the $\phi$-based zonal-flow energy~(\ref{eq:zonal_energy_balance}), governed by the competition between the $\bm{E}\times\bm{B}$ and magnetic-flutter stresses, in the main text, where $T_{\mathrm{exch}}$ is accordingly omitted from~(\ref{eq:zonal_energy_balance}). 

We do not, in this work, address the saturation of the microtearing modes that dominate the reverse-transition regime. For these modes, the zonal magnetic fields (zonal currents) can themselves contribute to saturation~\cite{Pueschel2020, Giacomin2023microtearing, Giacomin2024}; and indeed we see the size of the zonal parallel current become comparable (and exceed) the size of the zonal potential in the reverse transition [panels (j), (k), and (l) of Figure~\ref{fig:bpar_assessment}]. In the present zonal-flow energy budget, the zonal $A_\parallel$ enters only through the retained magnetic-flutter stress $\Pi_{A_\parallel}$.

\section{Geometric dependence of the linear drift term}
\label{app:linear_term}

Considering the radial component of the curvature drift present in Equation~(\ref{eq:Pi_lin_curv}) with an additional factor of $1 / B$ originating from the gyrofrequency, it is possible to show that in a toroidal system $\left \{\varrho, \theta, \zeta \right \}$ where $\varrho$ is some flux-surface label, $\theta$ is the poloidal angle and $\zeta$ is the toroidal angle, the curvature drive in Equation~(\ref{eq:Pi_lin_curv}) can be written as
\begin{equation}
    \frac{1}{B} \left(\mathbf{b} \times \left[ \left( \mathbf{b} \cdot \nabla \right)   \mathbf{b} \right] \right) \cdot \nabla \varrho = \frac{B_{\zeta}}{ 2 \mathcal{J}}  \frac{\partial}{\partial \theta}\left( \frac{1}{ B^2}  \right)
\label{eq:radial_curvature_drift}
\end{equation}
where $\mathcal{J}$ is the Jacobian, $B_{\zeta}$ is the current function and axisymmetry has been assumed. We note that when included within the flux-surface average of (\ref{eq:Pi_lin_curv}), the factor of the Jacobian that arises from the corresponding volume integral cancels with that in  (\ref{eq:radial_curvature_drift}). 

The magnitude of the magnetic field can be calculated using the form $\mathbf{B} = \nabla \zeta \times \nabla \psi  + B_{\zeta} \nabla \zeta$, such that $B^2 = \left( \left | \nabla \psi \right |^2 + B_{\zeta}^2 \right) / R^2$. Evaluating $\left | \nabla \psi \right |^2$, one can show \cite{Candy2009, dudding2022-2} that
\begin{equation}
    \left | \nabla \psi \right |^2 = {\psi^{\prime}}^{2} \left[ \left(\frac{\partial R}{\partial \theta}\right)^2 + \left(\frac{\partial Z}{\partial \theta}\right)^2\right]\left(\frac{\partial R}{\partial \varrho} \frac{\partial Z}{\partial \theta} - \frac{\partial R}{\partial \theta} \frac{\partial Z}{\partial \varrho} \right)^{-2}
\label{eq:grad_psi_squared}
\end{equation}
where $\psi^{\prime} = \mathrm{d} \psi / \mathrm{d} \varrho$. The current function can be written in terms of the safety factor via
\begin{equation}
    B_{\zeta} = q \psi^{\prime} \left[ \frac{1}{2 \pi}\int_{- \pi}^{\pi} \frac{1}{R} \left( \frac{\partial R}{\partial \varrho} \frac{\partial Z}{\partial \theta} - \frac{\partial R}{\partial \theta} \frac{\partial Z}{\partial \varrho} \right) \, \mathrm{d} \theta \right]^{-1}.
\label{eq:current_function}
\end{equation}
The variation in $B^2$ with poloidal angle evident in (\ref{eq:radial_curvature_drift}) therefore comes from two factors: $1/R^2$, attributable to both the poloidal and toroidal fields, as well as $\left | \nabla \psi \right |^2$, originating from the poloidal field only. By changing the geometry of flux surfaces, the poloidal variation of the magnetic field strength can be altered, affecting the linear stress term. To estimate the degree to which the linear stress is affected by such changes, we consider order of magnitude estimates of (\ref{eq:radial_curvature_drift}). Splitting the magnetic field strength into its poloidal and toroidal components via $B^2 = B_{\mathrm{pol}}^2 + B_{\mathrm{tor}}^2$ where $B_{\mathrm{pol}}^2 = \left | \nabla \psi \right |^2 / R^2$ and $B_{\mathrm{tor}}^2 = B_{\zeta}^2 / R^2$ we can write
\begin{equation}
    \frac{1}{B} \left(\mathbf{b} \times \left[ \left( \mathbf{b} \cdot \nabla \right)   \mathbf{b} \right] \right) \cdot \nabla \varrho = \frac{1}{ 2 \mathcal{J} B_{\zeta}}  \frac{\partial}{\partial \theta}\left( \frac{R^2}{ 1 + B_{\mathrm{pol}}^2 / B_{\mathrm{tor}}^2}  \right)
\label{eq:radial_curvature_drift_alt}
\end{equation}
such that for $B_{\mathrm{pol}}^2 / B_{\mathrm{tor}}^2 \ll 1$, the argument of the derivative with respect to poloidal angle is simply $R^2$. As the ratio increases to $B_{\mathrm{pol}}^2 / B_{\mathrm{tor}}^2 \lesssim 1$, this argument changes in a generally non-trivial manner. Using (\ref{eq:grad_psi_squared}) and (\ref{eq:current_function}), the field component ratio can be estimated as $B_{\mathrm{pol}}^2 / B_{\mathrm{tor}}^2 \sim \epsilon^2 / q^2$, where $\epsilon = r / R$ is the inverse aspect ratio of the surface. Thus, for large-aspect-ratio and/or large $q$, changes to the poloidal field through shaping are expected to minimally influence the form of the curvature drift. A similar estimate for the entire drift term yields
\begin{equation}
\begin{split}
    \frac{1}{B} \left(\mathbf{b} \times \left[ \left( \mathbf{b} \cdot \nabla \right)   \mathbf{b} \right] \right) \cdot \nabla \varrho & \sim \frac{1}{\psi^{\prime}}\frac{\epsilon}{q}
\end{split}
\label{eq:radial_curvature_drift_alt2}
\end{equation}
such that the magnitude of the linear stress term is expected to be small for $B_{\mathrm{pol}} / B_{\mathrm{tor}} \sim \epsilon /q \ll 1$.

\end{document}